\documentclass[aip,jcp,floatfix,amsmath,amssymb,preprint]{revtex4-2}

\usepackage{amsmath}
\usepackage{amssymb}
\usepackage{times}
\usepackage{booktabs}
\usepackage{hyperref}
\usepackage{mathptmx}
\usepackage[version=4]{mhchem}
\usepackage{lineno}
\usepackage{multirow}
\usepackage{float}
\usepackage{xr}
\usepackage{graphicx}
\usepackage{dcolumn}
\usepackage{bm}
\usepackage[utf8]{inputenc}
\usepackage[T1]{fontenc}
\usepackage{mathptmx}
\usepackage{etoolbox}
\usepackage{xcolor}

\usepackage[normalem]{ulem}

\AtBeginDocument{%
  \heavyrulewidth=.08em
  \lightrulewidth=.05em
  \cmidrulewidth=.03em
  \belowrulesep=.65ex
  \belowbottomsep=0pt
  \aboverulesep=.4ex
  \abovetopsep=0pt
  \cmidrulesep=\doublerulesep
  \cmidrulekern=.5em
  \defaultaddspace=.5em
}
\makeatletter
\def\@email#1#2{%
\endgroup
 \patchcmd{\titleblock@produce}
  {\frontmatter@RRAPformat}
  {\frontmatter@RRAPformat{\produce@RRAP{*#1\href{mailto:#2}{#2}}}\frontmatter@RRAPformat}
  {}{}
}%
\makeatother

\makeatletter
\AtBeginDocument{\let\LS@rot\@undefined}
\makeatother

\def\supplementfilename{mainSI}

\newcolumntype{d}[1]{D{.}{.}{#1}}

\newcommand{\angstrom}{\mbox{\normalfont\AA}}

\begin{document}


\title[]{Molecular dynamics of ice-active solutions at ice-water interfaces}
\author{Benjamin M. Harless}
\author{Jasmine K. Sindelar}
\author{J. Daniel Gezelter}
\email{gezelter@nd.edu}
\affiliation{251 Nieuwland Science Hall, Department of Chemistry and Biochemistry, University of Notre Dame, Notre Dame, Indiana 46556, USA}
\begin{abstract}
Small molecules that interact strongly with water were the subject of this molecular dynamics (MD) study. These solutes include a cryoprotectant (DMSO), a polyalcohol [\ce{CH2(OH)2}], carboxylic acid conjugates (\ce{HCOOH} and \ce{HCOONa}), an ammonium salt (\ce{NH4Cl}), and two alkyl halide salts (\ce{NaCl} and \ce{NaF}). MD simulations were carried out for bulk supercooled liquids and solutions in contact with ice. Solute and water hydrogen bonding, orientational and translational order, and hydrogen bond jump dynamics were compared in bulk and as a function of distance from the solute molecules. Reverse non-equilibrium molecular dynamics (RNEMD) simulations were used to determine interfacial widths, friction coefficients ($\kappa$) with ice, and solution phase viscosities ($\eta$).  
Ionic solutes were found to reduce orientational and translational ordering near the ice interfaces. However, in bulk liquids, we find a correlation between orientational ordering and the statistics of water hydrogen bonds -- a donor-acceptor imbalance in water has the greatest impact on ordering in the bulk liquids.
Although ionic solutions exhibited similar effects on the water structure, the effect on dynamics depends most directly on donor-acceptor imbalance. Solutes that are hydrogen bond acceptors were found to slow hydrogen bond lifetimes relative to hydrogen bond donors. We also observed a direct correlation between the liquid phase hydrogen-bond jump times and shear viscosity. Finally, of all the solutes studied, only DMSO and sodium formate exhibited increased friction at the ice-water interface. 
\end{abstract}
\maketitle

\section{Introduction}
Chemical methods for controlling the growth of ice crystals fall into a number of categories. Cryoprotectants can either inhibit crystal growth or ice recrystallization. Many molecules have been shown to act in this capacity, sometimes at very low concentrations. Cryoprotectants include biological antifreeze proteins,\cite{Duman:1991aa, Fletcher:2001aa, Mangiagalli:2017aa,Bar-Dolev:2016aa} some synthetic dyes,\cite{Drori:2016aa} poly(vinyl alcohol),\cite{Inada:2003aa,Congdon:2013aa,Weng:2018aa,Bachtiger:2021aa} some sugars,\cite{Capicciotti:2012aa,Volk:2006aa} dimethyl sulfoxide (DMSO),\cite{Lee:2022aa,Volk:2006aa} and some small polyalcohols, such as glycerol, ethylene glycol, and propylene glycol. Deicing solutions contain molecules in sufficient concentration to lower the melting point of the solution.\cite{Gruber:2023aa} The most common deicing agents in widespread use are common salts, including sodium and calcium chloride. There has been renewed interest in the development of environmentally benign non-chloride deicing agents that have a low chemical oxygen demand (COD) or biochemical oxygen demand (BOD).\cite{Sawyer:2003aa} Formate salts are candidates for environmentally-friendly road deicing agents, with significantly lower COD than acetate salts and glycerol.\cite{Bang:1998aa} Properties of interest in this study include how these molecules affect the ice-liquid friction coefficient and the viscosity of the liquid layer adjacent to ice. Molecules that are capable of seeding and promoting ice growth include biological ice nucleating particles,\cite{Polen:2016aa} salts such as silver iodide,\cite{Glatz:2016aa} and some minerals (e.g. Feldspar) whose ice-nucleating activity can be enhanced by the presence of ammonium chloride.\cite{Whale:2022aa,Blow:2024aa} 

\begin{figure}[]
    \centering
    \includegraphics[width=\linewidth]{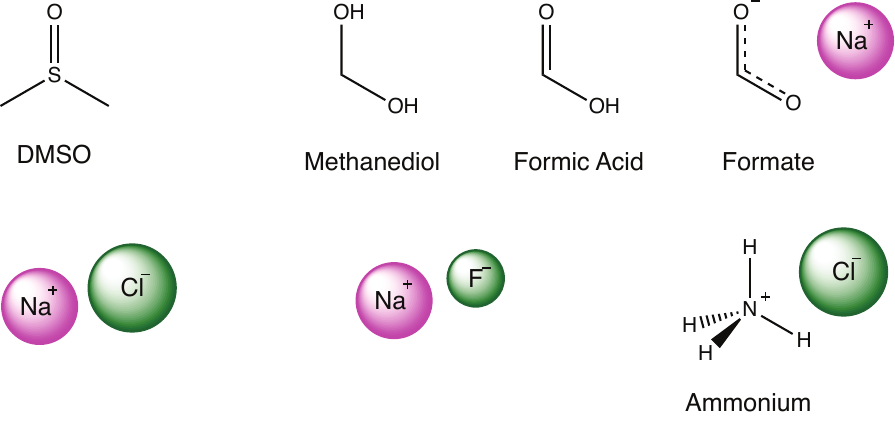}
    \caption{Solutes used in this study include a cryoprotectant  (DMSO), deicing agents (formate, NaCl), a promoter of ice nucleation (ammonium chloride), and a small polyalcohol (methanediol). Methanediol, formic acid, and formate differ in oxidation state (and protonation).  NaF is included to compare the effects of hydrogen bonding with NaCl.}
    \label{fig:MolecularStructures}
\end{figure}

    In this study, we focus on small-molecule cryoprotectants, deicing, and nucleation-enhancing solutes in supercooled liquid and ice-water interfacial solutions. Three of these, methanediol [\ce{CH2(OH)2}], formic acid (\ce{HCOOH}), and the formate anion (\ce{HCOO-}), comprise a family of molecules that differ only in protonation (see Fig. \ref{fig:MolecularStructures}). Formic acid mimics carboxylic acid groups present in proteins at low pH, while the formate anion mimics the same structures present at higher pH.  Formate salts are known deicing agents and are therefore directly relevant to this study. Methanediol is a simplified model for polyalcohols including PVA, glycerol,  ethylene glycol, and propylene glycol. The other solutes included in this study are DMSO, NaCl, NaF, and \ce{NH4Cl}. DMSO is a widely used cryoprotectant commonly used in plant vitrification solutions and medical tissue research.\cite{Lee:2022aa} NaCl is the most commonly used home deicing agent. While NaF shares most characteristics of NaCl, it can participate directly in the hydrogen bonding network of water. \ce{NH4Cl} has been shown to enhance ice nucleation\cite{Whale:2022aa,Blow:2024aa} and is the sole excess hydrogen bond donor in this study. 

    Understanding how these solutes change the structure and dynamics of surrounding liquid water may help to elucidate mechanisms behind ice inhibition and growth. Previous studies by Fitzner \textit{et al.}\cite{Fitzner:2019aa} and Lupi \textit{et al.}\cite{Lupi:2017aa} emphasized the importance of slow dynamics and disorder in the nucleation process. Others have worked on characterizing the impact of solutes featured in this study. Previous studies on DMSO by Lee and Baiz,\cite{Lee:2022aa} found significant disruption of the hydrogen bond network within the first solvation shell. Work by Whale\cite{Whale:2022aa} and Blow \textit{et al.}\cite{Blow:2024aa} has shown that \ce{NH4Cl} induces disorder of the water hydrogen bonding network, enhancing nucleation.
    
    In order to characterize the structure and dynamics of water molecules, we utilize three well-known order parameters: the tetrahedral order parameter,\cite{Chau:1998aa,Errington:2001aa} a translational order parameter,\cite{Truskett:2000aa} and hydrogen bond jump times.\cite{Laage:2006aa}  By calculating these order parameters with spatial resolution around the solute molecules, we investigate how ``ice-like'' the solution behaves near the solutes or ice interface. The analysis of the hydrogen bond jump times reveals how the solutes affect the hydrogen bond jump activation energy, shear viscosity and the interfacial friction between the ice sheet and liquid layer. We utilize the TIP4P-Ice model\cite{Abascal:2005aa} for water in all of the simulations described below. TIP4P-Ice reproduces the melting point of water, and the thermodynamic behavior, shear viscosity, and diffusion have been thoroughly investigated by Baran, \textit{et al.} over a wide temperature range.\cite{Baran:2023aa}  In sections II-IV, we provide details on the simulation and analysis methodologies and draw conclusions about connections between structural and dynamic properties of  cryoprotectant, deicing, and ice nucleation enhancing solutions.
    
\begin{figure}[]
    \centering
    \includegraphics[width=\linewidth]{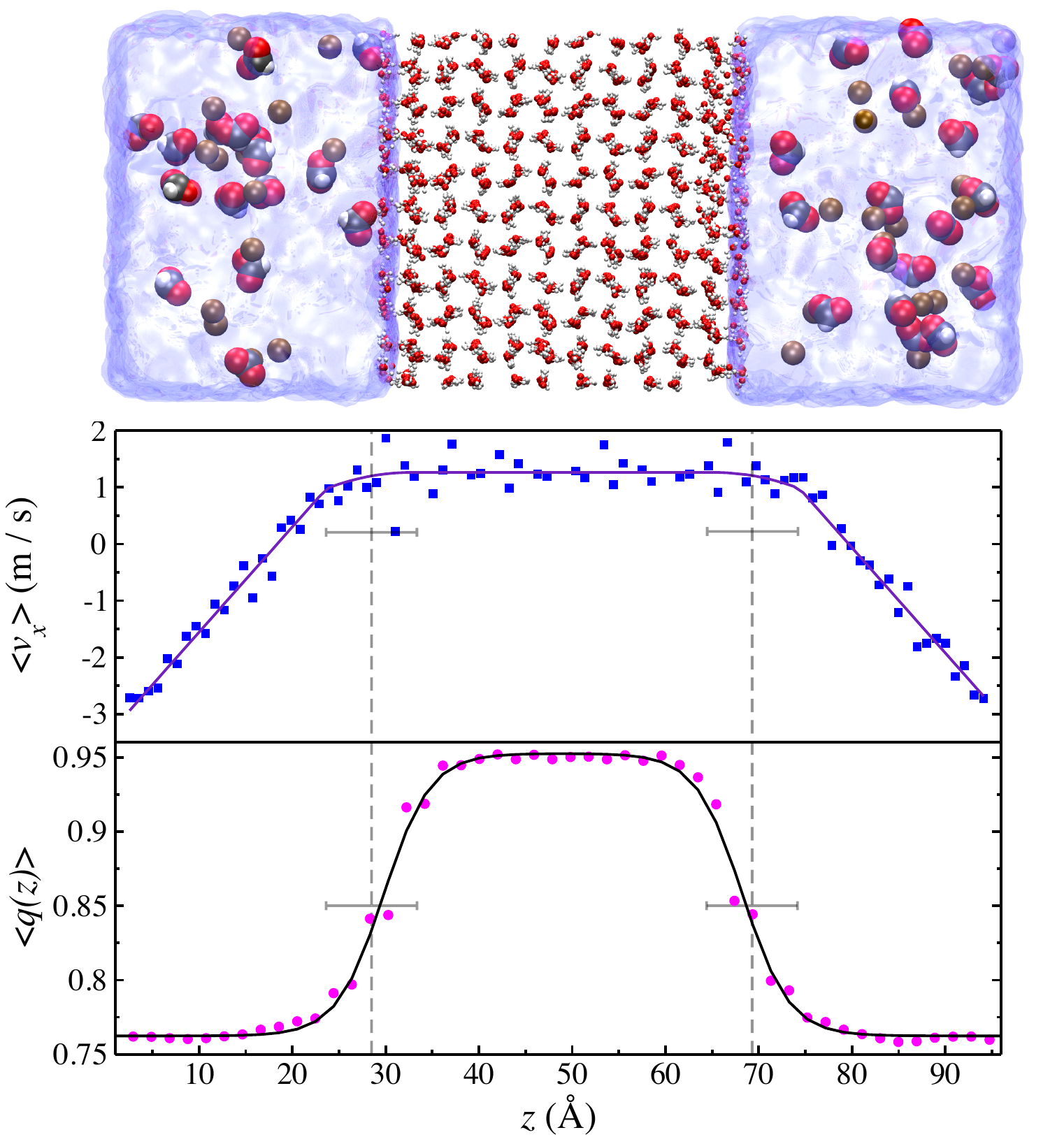}
    \caption{Top: A 0.5M sodium formate solution in contact with the basal facet of an ice-Ih crystal. Solute molecules are indicated with space-filling atoms, while the molecules in the crystal are shown in a ball and stick representation. Middle: the transverse velocity profile. Bottom: the tetrahedrality profile. Both are averages from the same five RNEMD simulations. The solid-liquid Gibbs dividing surfaces are indicated with gray dashed lines, and the $w_{10-90}$ that was used to find the solid-liquid friction coefficient $(\kappa)$ is indicated with horizontal bars. }
    \label{fig:SoluteIceSim}
\end{figure}

\section{Methodology}
\subsection{Force fields and models}
All molecular dynamics (MD) simulations were carried out using OpenMD, \cite{Drisko:2024aa} using the real space damped shifted force (DSF)\cite{Fennell:2006aa} model for electrostatic interactions with a cutoff of 12 \AA~ and a damping parameter $\alpha = 0.18 \angstrom^{-1}$. All simulations were done in simulation cells with periodic boundary conditions, and dimensions are chosen to give bulk-phase densities of $1 \mathrm{g~cm}^{-3}$ and solution phase solute concentrations of 1 M (or 0.5 M). 1~fs time steps were used for all solutes except for \ce{NH4Cl} which was simulated with a 0.5~fs time step.

Force field parameters for the \ce{CH2(OH)2}, \ce{CHOOH}, and \ce{HCOO-} anion species were adapted from the Generalized Amber Force Field (GAFF)\cite{Wang:2004aa} with charges on the \ce{CH2(OH)2} adapted from the work of Fennell, Wymer, and Mobley, which provides accurate descriptions of solvation free energies for alcohols.\cite{Fennell:2014aa}  DMSO was modeled using the  CHARMM General Force Field (CGenFF) parameters provided by Strader and Feller~\cite{Strader:2002aa} which have been previously shown to give accurate IR spectra when solvated in water.\cite{Lee:2022aa}  The ammonium cation parameters were adapted from the work of Kashefolgheta and Vila Verde, \cite{Kashefolgheta:2017aa} while the monatomic ions were modeled using the general 12-6 Lennard-Jones parameters from Li, Song, and Merz.\cite{Li:2015aa}
In all cases, the molecular and ionic species were solvated in either liquid water boxes or interfacial ice/water systems composed of TIP4P-Ice rigid body models for water,\cite{Abascal:2005aa} which have been used previously in interfacial studies by others\cite{Benet:2014aa}.  Force field parameters for all species are provided in the supplementary material (SM).

\subsection{Simulation protocol}
\subsubsection{Supercooled Liquids}
    For the aqueous DMSO, \ce{HCOONa}, \ce{HCOOH}, \ce{NaCl}, \ce{NaF}, \ce{CH2(OH)2}, and \ce{NH4Cl} solutions, five independent configurations for each solution were created using packmol with unique random seeds.\cite{Martinez:2009aa} Solute counts, water counts, solution densities, simulation dimensions, and solute concentrations are provided in Table \ref{tab:1MSolnParam}. Five independent configurations of neat liquid water were created for a negative control. Initial velocities were sampled from a Maxwell-Boltzmann distribution corresponding to temperatures from 250 to 300~K in 10~K increments. Equilibration was carried out in the canonical (NVT) ensemble  to keep solution concentrations and densities constant at different temperatures.

    Equilibration to reach a fixed temperature was done over 210 ps, with data collection for analysis of structural order parameters carried out over a 1 ns period, with configurations stored every ps. Sampling for dynamic properties (time correlation functions and hydrogen bond jump times) was done over an additional ns of simulation in the microcanonical (NVE) ensemble, with configurations sampled every ps. The total energy for these simulations was set to the mean energy from the canonical simulations.

    The end points of these equilibration simulations were used as the starting configurations for reverse non-equilibrium molecular dynamics (RNEMD) simulations  in a manner similar to our previous work.\cite{Louden:2017aa} First, the end points of these systems were replicated twice in the $z$ direction to have a comparable $z$ length to our interfacial systems. A momentum flux, $j_z(p_x) = 1 \times10^{-6}$ amu \AA$^{-2}$ fs$^{-1}$, normal to the $z$ axis was imposed using velocity shearing and scaling (VSS) every 2 fs over the course of a nanosecond as the system approached steady state. Once the ice-liquid interfaces had reached steady-state transverse velocities, data collection was performed using the same simulation parameters. Velocity, temperature, and density profiles were sampled every 2 fs.

\begin{table}
\bibpunct{}{}{,}{n}{,}{,}
    \centering
    \caption{Densities, concentrations, and molecular compositions of supercooled liquid simulations}
\label{tab:1MSolnParam}
\begin{tabular}{ l | c c c c c c c}
                \midrule
 
Solution & $N_{\mathrm{solute}}$ & $N_{\mathrm{ions}}$ & $N_{\ce{H2O}}$ & $L_{\{x,y,z\}}$ (\AA) & $\rho (\mathrm{~g~cm^{-3}})$ & [] (M) \\
\midrule
1M DMSO & 17 & - & 870 & 30.449 & 1.000 & 1.000\\
1M \ce{CH2(OH)2} & 17 & - & 898 & 30.449 & 0.999 & 1.000\\
1M CHOOH & 17 & - & 900 & 30.449 & 0.999 & 1.000\\
1M \ce{HCOONa} & - & 34 & 879 & 30.449 & 0.999 & 1.000\\
0.5M \ce{HCOONa} & - & 18 & 965 & 31.0345 & 1.000 & 0.500\\
1M NaCl & - & 34 & 889 & 30.449 & 1.000 & 1.000\\
1M NaF & - & 34 & 904 & 30.449 & 1.000 & 1.000\\
1M \ce{NH4Cl} & - & 34 & 893 & 30.449 & 1.000 & 1.000\\
0.5M \ce{NH4Cl} & - & 18 & 972 & 31.0345 & 1.000 & 0.500 \\
\bottomrule
\end{tabular}
\end{table}

\subsubsection{Ice interfaces with de-icing solutions}
    To study the effects of the solutes on the ice--water interface, 1M (and a few 0.5M) solutions were simulated in contact with the basal faces of an ice Ih crystal. To construct interfacial systems, packmol~\cite{Martinez:2009aa} was used to create five independent configurations for each solute.  The solute molecules were randomly placed in regions corresponding to the liquid portion of the simulation. These were then combined with a low energy ice--water interface generated by Louden and Gezelter.\cite{Louden:2017aa} All ice--water interfaces studied here were basal $\{0001\}$ interfaces.  Any water molecules found within 2.2 \AA~ of solute atoms were removed, resulting in a small variation in the number of water molecules. Solute counts, average water molecule counts, concentrations and box geometries are given in Table \ref{tab:IceParam}.  A more detailed description of how the original interfaces were constructed from orthorhombic starting structures~\cite{Hirsch:2004aa} is given in the supplementary material and Refs. \citenum{Louden:2013aa} and \citenum{Louden:2017aa}.

    Initial velocities were sampled from a Maxwell--Boltzmann distribution corresponding to 50~K. As in the supercooled liquid simulations, equilibration was carried out in the canonical (NVT) ensemble. Initial equilibration at a low temperature allowed the system to relax from the initial introduction of the solutes, and prevented premature disruption of the ice crystal. The simulations were then brought up to final temperatures of 230--270 K in 10 K increments using resampling from Maxwell--Boltzmann velocity distributions. In order to prevent premature melting of the ice sheets, resampling and relaxation at these fixed temperatures were done over relatively short periods: 20 ps (for most systems), 50 ps (for NaCl and NaF), or 200 ps (for 1M \ce{NH4Cl}). For these salts, the initial placement of the ionic species in the solution phase resulted in relatively high energy configurations, and additional time was needed to help the local water structure relax around these ions. Data collection was conducted in the same manner as the supercooled solutions.

    The end points of the 270 K simulations were used as the starting configurations for RNEMD simulations. In addition to the applied momentum flux, $j_z(p_x) = 1.0 \times 10^{-6}$ amu \AA$^{-2}$ fs$^{-1}$, a weak thermal flux normal to the ice surface, $J_q = -2.0 \times 10^{-6}$ kcal mol$^{-1}$ \AA$^{-2}$ fs$^{-1}$, was also imposed. The momentum and thermal fluxes were applied using velocity shearing and scaling (VSS-RNEMD), and the systems approached steady state conditions over the course of a nanosecond. The imposed momentum flux results in the formation of a velocity gradient in bulk liquids, but frictional heating can also raise the temperature at the interface. For this reason, an opposing thermal flux was required to keep the ice / solution interface at 270 K while the whole system evolved under shearing conditions. In \ce{NH4Cl}, the high frequency N--H bonds required a lower thermal flux, $J_q = -4.0 \times 10^{-7}$ kcal mol$^{-1}$ \AA$^{-2}$.

    Once the ice--liquid interfaces had reached steady-state transverse velocities, data collection was performed using the same simulation parameters. Configurations were stored every 100 fs over an additional ns, although velocity, temperature, and density profiles were sampled every 2 fs. 

\begin{table}
    \centering
    \caption{Composition of ice interface simulations. All simulation cells were $46.74 \angstrom \times 38.83 \angstrom \times 97.7 \angstrom$, and  contain a 35.9 \angstrom~slab of ice-Ih presenting two basal $\left\{0001\right\}$ facets to the solution, as seen in Fig. \ref{fig:SoluteIceSim}. Only the liquid region was used to calculate solute concentrations.}
\label{tab:IceParam}
\begin{tabular}{ l | c c c c c c }
                \toprule
Solution & $N_{\mathrm{solute}}$ & $N_{\mathrm{ions}}$ &$\langle N_{\ce{H2O}}\rangle$ (liq) & $N_{\ce{H2O}}$ (ice) & Solution [] (M)\\
\midrule
Pristine & - & - & 3692 & 1800   & 0\\
1M DMSO & 67 & - & 3278 & 1800 &  0.992\\
1M \ce{CH2(OH)2} & 67 & - & 3407& 1800  & 0.992\\
1M CHOOH & 67 & - & 3461& 1800 & 0.992\\
1M \ce{HCOONa} & - & 134 & 3427& 1800   & 0.992\\
0.5M \ce{HCOONa} & - & 68 & 3544& 1800  & 0.503\\
1M NaCl & - & 134 & 3535& 1800  &  0.992\\
1M NaF & - & 134 & 3516& 1800  &  0.992\\
1M \ce{NH4Cl} & - & 134 & 3428& 1800 & 0.992 \\
0.5M \ce{NH4Cl} & - & 68 & 3557 & 1800 & 0.503 \\
\bottomrule
\end{tabular}
\end{table}

\section{Results}

\subsection{Hydrogen bonding effects due to the presence of solute molecules}
Solutes that participate in the water hydrogen bonding network can do so as equitable participants or as excess donors or acceptors. This behavior is a distinguishing characteristic of the small molecules used in this study. For example, \ce{CH2(OH)2} and \ce{HCOOH} can participate more fully in the water hydrogen bonding network, while DMSO, \ce{HCOONa}, \ce{NaF}, and \ce{NH4Cl} all create distinct donor-to-acceptor imbalances in the solution. \ce{NaCl} can disrupt the water hydrogen bond network without participating in the network itself.

We determine if a hydrogen bond has been formed between two molecules using the geometric criteria of Luzar and Chandler.\cite{Luzar:1996aa}  We identify a hydrogen bond between two molecules if the distance between their potential donor (D = O--H, N--H, F--H) and acceptor (A = O, N, F) sites, $r_\mathrm{DA} < 3.5$~\AA and the DHA bond angle, $\theta_\mathrm{DHA} < 30^\circ$.  The mean hydrogen bonding statistics are provided in Table \ref{tab:HbondAcceptorDonor}.  Note that we observe a weak dependence on temperature for average hydrogen bonding in these solutions, but the difference between acceptors and donors is independent of temperature. Temperature-dependence of this data is provided in the supplementary material.

\begin{table}
    \centering
    \caption{Average number of hydrogen bonds accepted or donated by solutes. Solute acceptor--donor mismatch creates a mismatch in the surrounding water. Uncertainties in the last digit are shown in parentheses. All concentrations are 1 M unless stated otherwise.}
\label{tab:HbondAcceptorDonor}
\begin{tabular}{ l | c c |c }
                \toprule
Solution & $\langle N^\mathrm{solute}_\mathrm{acceptor} \rangle$ & $\langle N^\mathrm{solute}_\mathrm{donor} \rangle$ & 
$\langle N^{\ce{H2O}}_\mathrm{acceptor} \rangle - \langle N^{\ce{H2O}}_\mathrm{donor} \rangle$\\
\midrule
DMSO & 2.37(1) & - & -0.046(1) \\
\ce{CH2(OH)2} & 2.80(1)  & 1.91(1)  & -0.017(1) \\
CHOOH & 1.63(2) & 0.97(1)  & -0.012(1) \\
\ce{HCOONa} & 5.99(2) & - & -0.116(1) \\
0.5M \ce{HCOONa} & 5.87(6) & - & -0.055(2) \\
NaCl & - & - & 0 \\
NaF & 5.84(4)  & - & -0.110(2) \\
\ce{NH4Cl} & - & 3.13(2) & 0.060(1) \\
0.5M \ce{NH4Cl} & - & 3.22(1) & 0.030(1) \\
\bottomrule
\end{tabular}
\end{table}

\subsubsection{Local tetrahedral ordering}
    The tetrahedral order parameter is a structural measure of the local orientational ordering of water molecules. This parameter describes the distribution of angles between oxygen atoms in water and other heavy (non-hydrogen) atoms in the first solvation shell. The order parameter was originally formulated by Chau and Hardwick\cite{Chau:1998aa} and was developed more fully for water by Errington and Debenedetti.\cite{Errington:2001aa} It has been widely used in previous simulations of water and ice, and was renormalized to account for under- or over-coordinated molecules and water molecules at interfaces.\cite{Louden:2013aa,Louden:2017aa} The local tetrahedral order parameter for molecule $i$,
\begin{equation}
q_i=\left (1 - \frac{9}{2n(n-1)}\sum_{j=1}^{n-1}\sum_{k=j+1}^{n} \left (cos\psi_{jik} + \frac{1}{3} \right) ^2 \right) ~,
\label{eq:qi}
\end{equation}
where $\psi_{jik}$ is the angle formed between three adjacent heavy atoms $(j,i,k)$, where $i$ is the central atom and the heavy atoms can come from either water or solute molecules. The double sum accounts for $n$ neighboring heavy atoms within a cutoff radius ($r_c < 3.5$ \angstrom). In this formulation, $n$ is not required to be the four closest neighbors, so this form is appropriate for situations involving either under- or over-coordination in the first solvation shell.\cite{Louden:2017aa}
The normalization adjusts the number of neighbors around the central atom, and evaluates to 3/8 for molecules with four neighbors.\cite{Errington:2001aa} $q_i$ ranges from 0, describing an isolated molecule, to 1, in a perfectly tetrahedral network around $i$ (e.g., ice crystals).

\begin{figure}[]
   \centering
    \includegraphics[width=\linewidth]{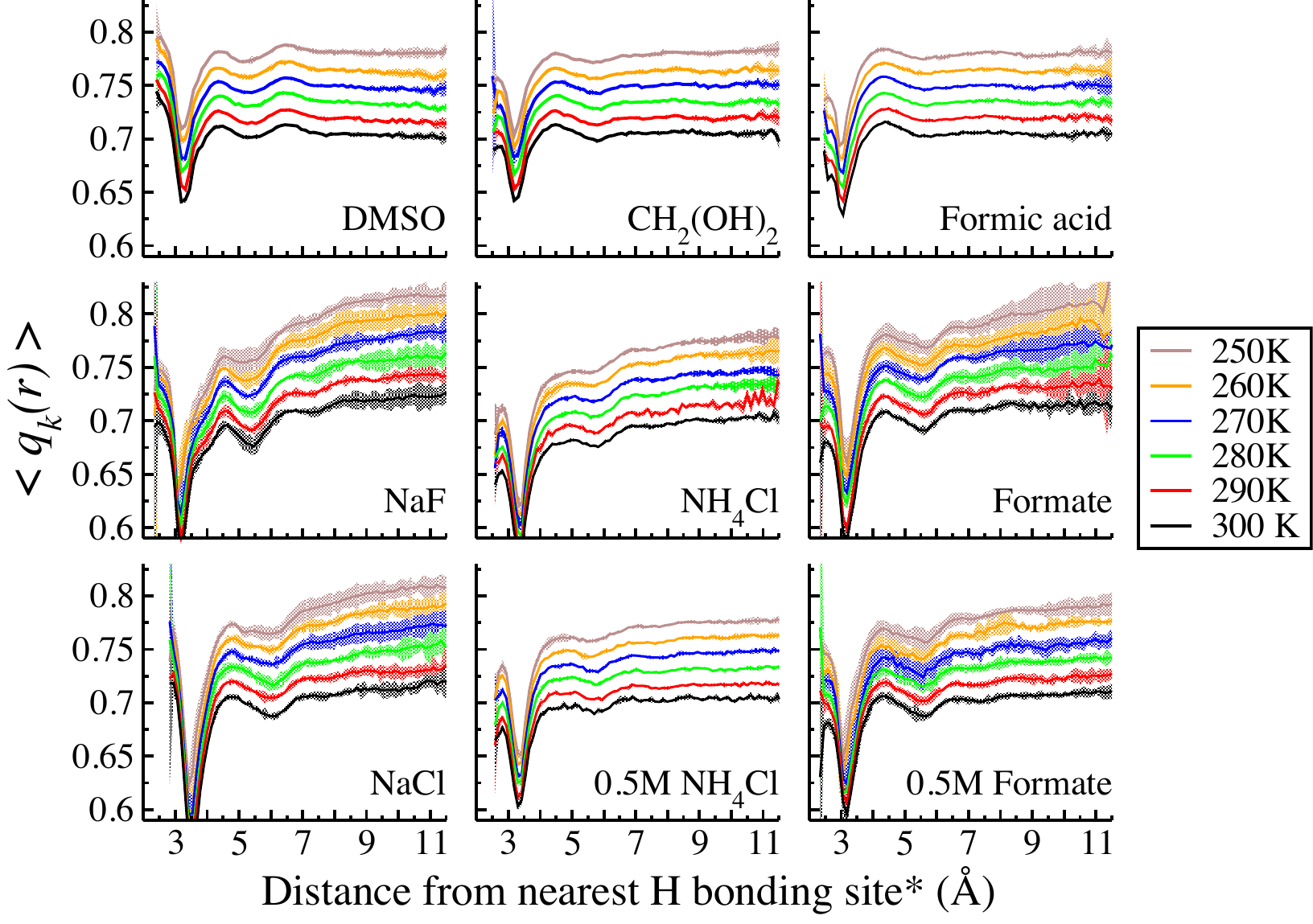}
   \caption{Tetrahedral order, $q_k(r)$, as a function of distance from the nearest solute hydrogen bonding site (\ce{Cl-} for NaCl). All species disorder nearby water layers, but differences appear at larger distances. Notably, DMSO increases order in the water in intermediate distances, while the effects of NaF, NaCl, and formate extend to longer distances.}
   \label{fig:tetrahedrality by r data}
\end{figure}

Because $q_i$ measures the local environment around molecule $i$, it can be averaged over a collection of molecules or queried as a function of molecular distance $r$ from another site $k$,
\begin{equation}
q_k(r)= \int_0^\infty \frac{1}{N_i(r)} \sum_{i}^{N_i(r)} q_i \delta(r_{ik} - r) dr~,
\label{eq:qr}
\end{equation}
where $N_i(r)$ is the total number of atoms of type $i$ at distance $r$ and $r_{ik}$ is the distance between $i$ and $k$.
The dependence of $q_k(r)$ on the distance from the small molecule inclusions indicates whether the presence of these molecules disrupts (or enhances) the ordering of the nearby water environment. See, for example, the distance dependence of $q_k(r)$ for water molecules in the bulk supercooled liquids in Fig. \ref{fig:tetrahedrality by r data}.

Although disordering is seen close to a molecular inclusion, the overall solution phase behavior can actually order the solvent relative to the neat water solution seen in Figs. \ref{fig:QkTkTau270} and S1 in the supplementary material. Counter-ions introduce noise at longer distances relative to their molecular counterparts, although this effect is small in \ce{NH4Cl}. All molecular solvents have an ordering effect relative to the bulk with DMSO impacting solvent out to $\sim$ 0.6 nm. The deicers appear to increase order at distances up to $\sim$~1 nm.

\begin{figure}[]
    \centering
    \includegraphics[width=0.6\linewidth]{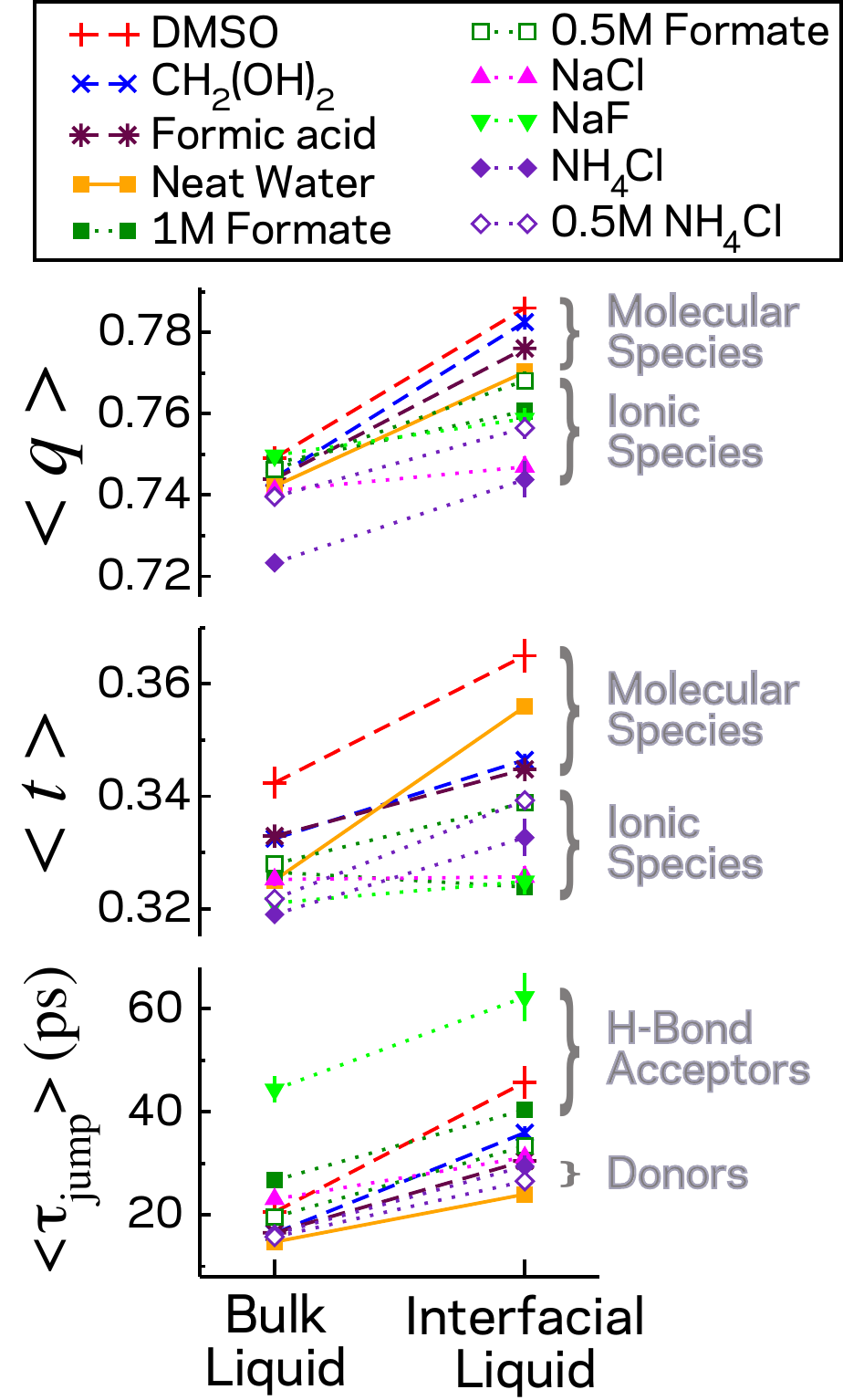}
    \caption{Tetrahedrality $\langle q \rangle$, translational order $\langle t \rangle$, and hydrogen bond jump times $\langle \tau_{jump} \rangle$ at ~270 K for bulk supercooled liquids and  interfacial solutions adjacent to ice.  Molecular solutes generally order the interface, while the ionic solutes disorder the interfacial liquid relative to pure water.}
    \label{fig:QkTkTau270}
\end{figure}

The local tetrahedrality can be sampled along a spatial axis of the simulation cell, notably along the $z$ coordinate, perpendicular to the ice face, with $N_z$ atoms in each bin,
\begin{equation}
q_i(z)= \int_{0}^{L_z} \frac{1}{N_{z}} \sum_{i=1}^{N_i} q_i  \delta(z_{i} - z) dz ~.
\label{eq:qz}
\end{equation}
In most ice/water interfaces, this parameter varies smoothly from a liquid state value $(q_i \sim 0.75)$ to ice-like ordering $(q_i \sim 0.95)$ inside the ice slab.  Fitting the spatial dependence of a general order parameter $O(z)$ in these simulations is carried out using a generic hyperbolic tangent function,
\begin{equation}\label{eq:tanH}
    O(z) = O_\mathrm{liq} + \frac{O_\mathrm{ice} - O_\mathrm{liq}}{2} \left( \tanh\left({\frac{z - l}{w}}\right) - \tanh\left({\frac{z-r}{w}}\right) \right)~,
\end{equation}
where $l$ and $r$ are the locations of the Gibbs dividing surfaces between water and ice and $w$ is the width of the ice/water interface for that order parameter. Haymet \textit{et al.} used a similar measure, the ``10--90'' width of an interface, to estimate the distance over which an order parameter dropped from 90\% $\rightarrow$ 10\% of the difference between the solid and liquid values.  We note that the Haymet 10--90 width is a scalar factor of the hyperbolic tangent width, $w_{10-90} = 2.197 w$. 

Interfacial widths using the tetrahedral and translational order, in addition to the dynamic width computed via the hydrogen bond jump times, are given in Fig. \ref{fig:interfacialwidthdata}. Widths were averaged across the 230--270~K temperature range, as there was no significant temperature dependence. Average widths for the different temperatures are provided separately in the supplementary material.

The solute dependence of tetrahedral ordering behavior correlates with the magnitude of the hydrogen bonding effects in Table \ref{tab:HbondAcceptorDonor}. The excess donor of hydrogen bonds (\ce{NH4Cl}) significantly disrupts the solution structure. \ce{NaCl} behaves similarly to bulk water, \ce{HCOOH}, and \ce{CH2(OH)2}. Both \ce{HCOONa} solutions behave similarly, showing no concentration dependence, and both produce increasing tetrahedral order. DMSO and NaF (both excess hydrogen bond acceptors)  have only a small impact on solution phase ordering.

Orientational order in the confined liquid in the ice--water simulations is increased relative to the supercooled systems, independent of solute. All studied molecular species increase the orientational order of the interfacial liquid relative to the pristine ice--water system. The ionic species all exhibit more orientational disorder relative to the pristine ice--water system. Average bulk tetrahedrality across different temperatures is given in Fig. S1.

\begin{figure}[]
    \centering
    \includegraphics[width=\linewidth]{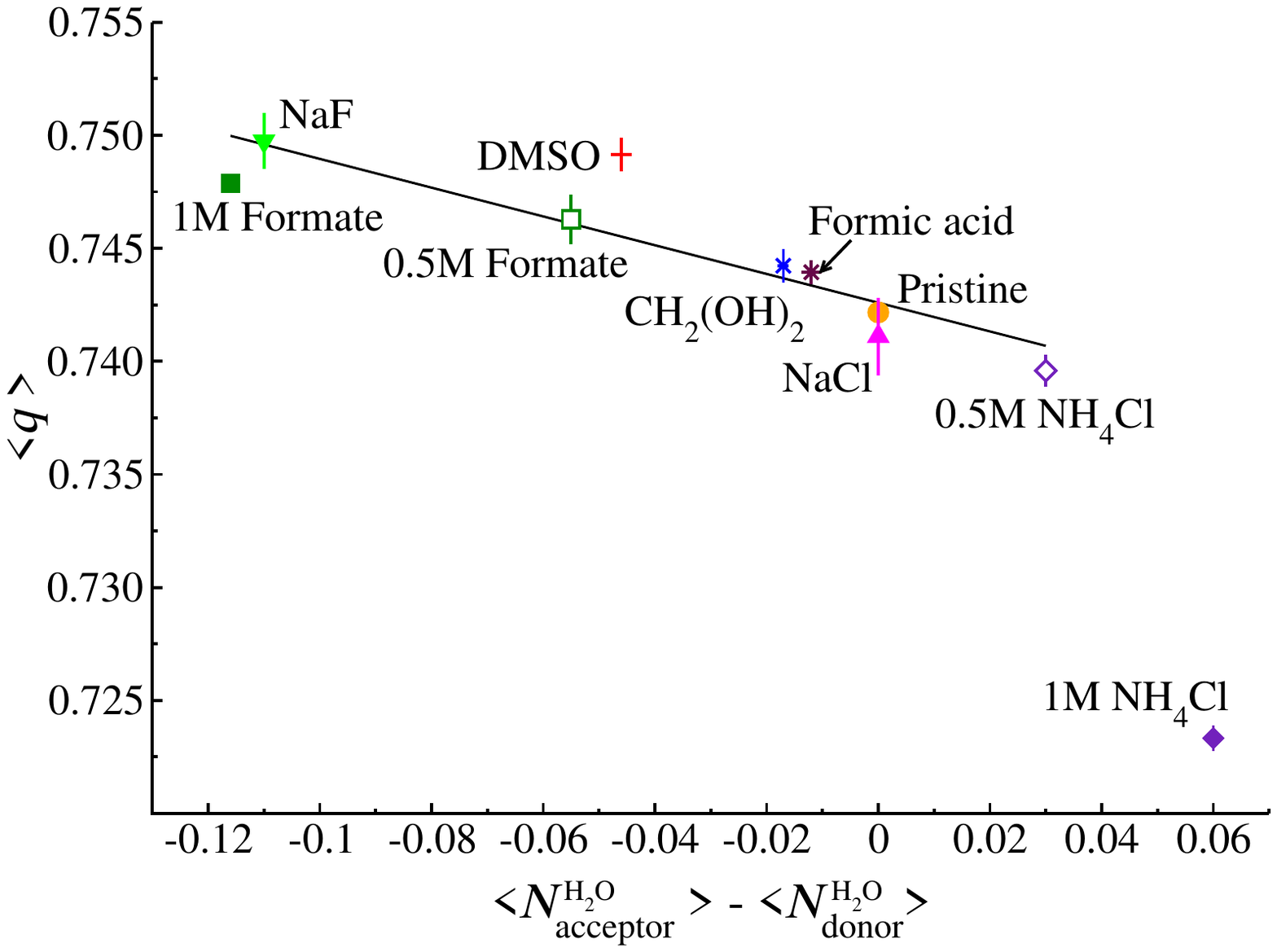}
    \caption{Tetrahedrality in supercooled liquid solutions (at 270K) correlates strongly with the overall balance of water hydrogen bonds.  Solutes that are excess acceptors create order, while excess donors reduce orientational ordering.}
    \label{fig:HBondvsQk}
\end{figure}

\subsubsection{Local translational ordering}

\begin{figure}[]
    \centering
    \includegraphics[width=\linewidth]{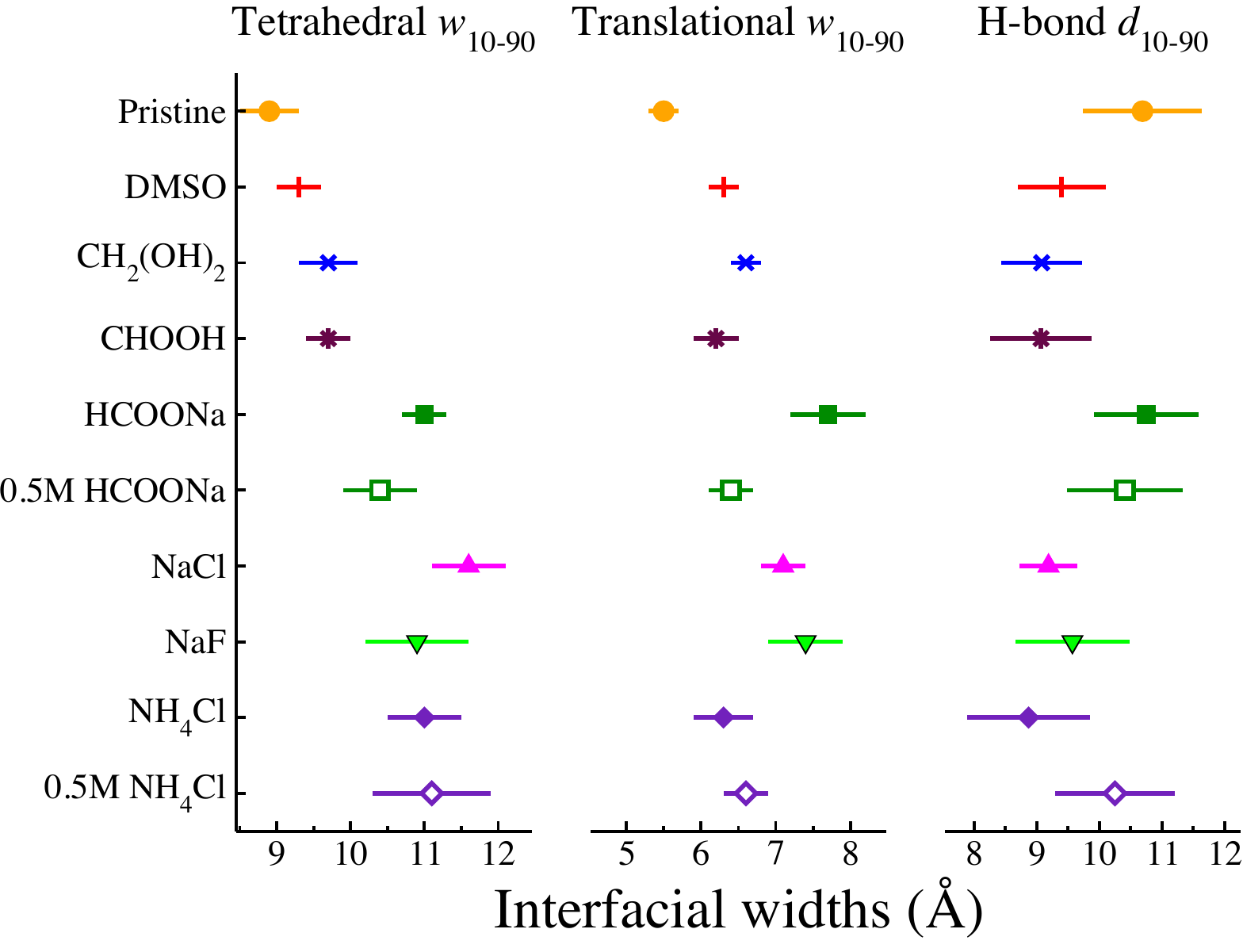}
    \caption{Ice interfacial widths for the basal facet in contact with small molecule solutions. These widths are based on tetrahedral and translational order parameters as well as the hydrogen bond jump times. Solutes generally increase the structural width of the interface relative to the pristine systems, and ionic solutions have a larger impact than the molecular species. 95\% confidence intervals in the widths are indicated with horizontal bars.}
    \label{fig:interfacialwidthdata}
\end{figure}

The translational order parameter is another measure of the local ordering of surrounding molecules around a central molecule. This order parameter takes advantage of the differences in structure in the pair correlation function, $g(r)$, specifically for oxygen--oxygen pairs in water environments,\cite{Truskett:2000aa,Shell:2002aa}
\begin{equation}
t =  \frac{1}{\xi_c}\int_0^{\xi_c}  |g(\xi) - 1)| d\xi
\end{equation}
Here $\xi=r\rho^{\frac{1}{3}}$ where $\rho$ is the number density of the species of interest (and $\xi_c=r_c\rho^{\frac{1}{3}}$ at the cutoff radius).  Larger values of $t$ indicate greater degrees of translational order around the central molecule.  As with orientational ordering, the translational order parameter can also be sampled spatially as a function of the axis perpendicular to the ice surface. Ice and interfacial liquid translational order parameters were computed using the same method described in Equation \ref{eq:tanH}. Estimated widths of the ice/water interfaces based on local translational ordering are given in Fig. \ref{fig:interfacialwidthdata}.

The molecular solutes increase the translational order, with DMSO having the largest effect. DMSO was the only solute that ordered the interfacial liquid more than the neat ice--water system. The ionic solutes decreased order, with the deicers creating the largest disruption in the interfacial solution. The temperature dependence of the supercooled solution, interfacial systems, and ice crystals is given in Fig. S2 in the supplementary material.

All solutes increase both the orientational and translational widths of the interface relative to the pristine ice--water interface. Additionally, the widths of the interfaces measured using tetrahedrality are wider ($\sim$~3 \AA) than the same width captured using translational order parameters.  In general, the presence of ionic species has a larger impact on tetrahedral width than on translational width although the difference is only $2-3$ \AA.

\subsection{Solute rejection from the interfacial region}
Previous simulations by Berrens \textit{et al.} have indicated that salt (NaCl) can incorporate into the quasi-liquid layer on ice,\cite{Berrens:2022aa} while x-ray diffraction experiments by Tsironi \textit{et al.} indicate that uptake into growing ice crystals occurs at reduced concentrations relative to the surrounding brine liquid phase\cite{Tsironi:2020aa}. Recent experiments by Sivells \textit{et al.} have shown that there is a higher propensity for the uptake of anions relative to cations, although the salt uptake remains at relatively low 1--20 mM concentrations when the surrounding salt solutions are at 100 mM.\cite{10.1063/5.0141057}  

\begin{figure}[]
    \centering
    \includegraphics[width=\linewidth]{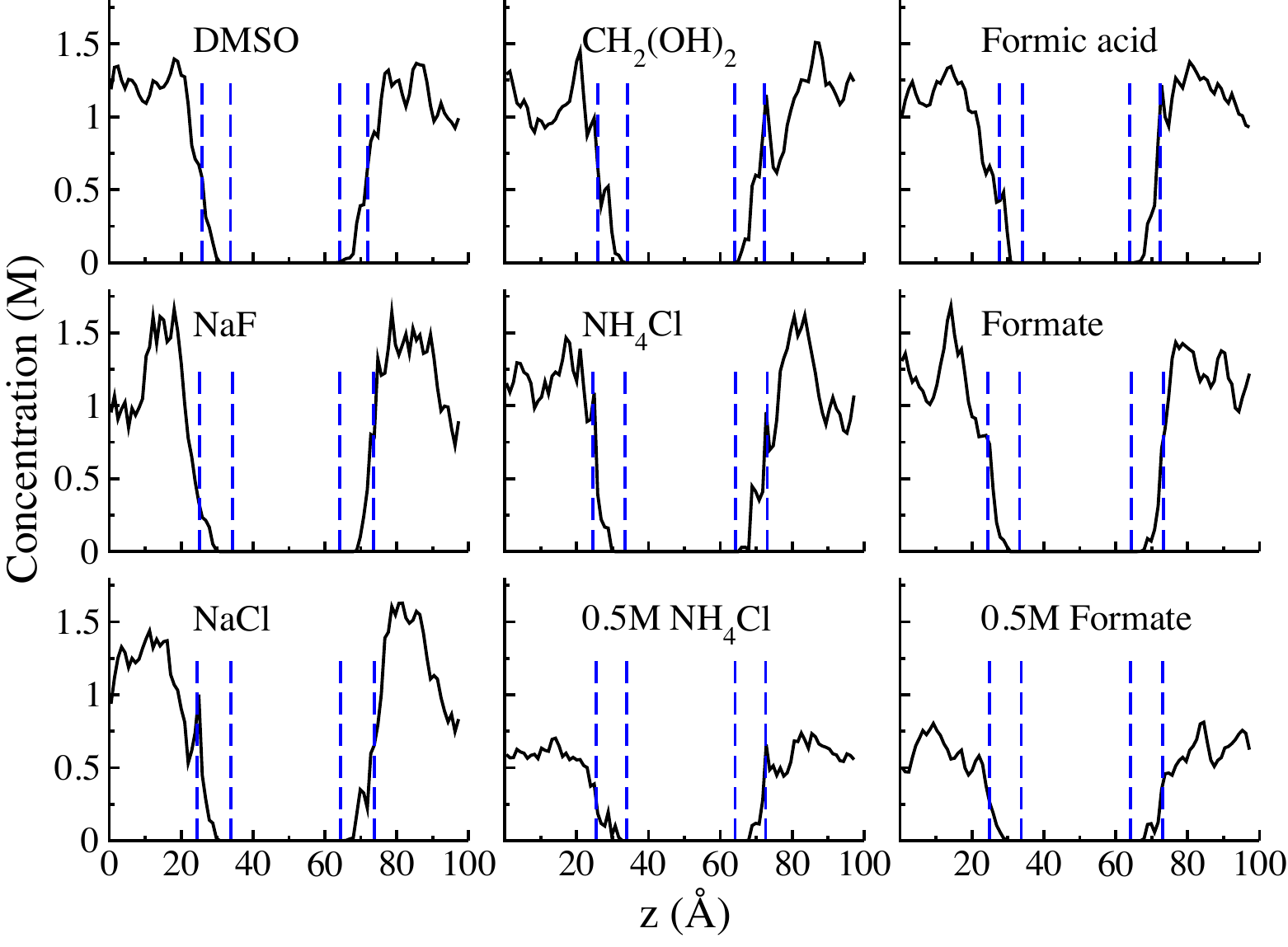}
    \caption{Average local solute concentrations as a function of location in the simulation cell. The ice regions in the center of each plot are bounded by four lines corresponding to the 90\% (inner) and 10\% (outer) bounds of the orientational order parameter, $q(z)$. All of the solutes exhibit rejection from the interfacial layer.}
    \label{fig:uptake}
\end{figure}

For the solutes in this study, fig. \ref{fig:uptake} shows the local solute concentrations as a function of location in the simulation cell during the shearing (RNEMD) simulations. All of the solutes exhibited rejection from the interfacial layer (relative to their bulk concentration), in agreement with previous experimental studies on brine rejection. We do not observe any differences in anion vs cation uptake into the interfacial region (see supplementary material), nor do we observe solute inclusion into the crystalline regions inside the 90\% $q(z)$ boundary defining ``bulk'' ice.

We note that the interfacial layer in these simulations is not directly analogous to the QLL that develops on ice / vapor interfaces, and that very little growth of the ice layer can take place during transverse shearing in these simulations. These differences may help explain the relatively small degree of observed solute uptake.

Integrating the solute concentration between the 10\% and 90\% boundaries of the orientational order parameter helps quantify the degree of solute uptake (see table \ref{tab:uptake}). We note that the uptake is concentration-dependent for the species simulated at multiple concentrations. However, we are unable to discern which of the solutes will permeate the interfacial region most readily as all of the uptake concentrations fall within the 95\% confidence intervals of the other 1M solute solutions. The uptake concentrations fall between 14\%--27\% of the solution phase concentrations, in general agreement with the experiments of Sivells \textit{et al.}\cite{10.1063/5.0141057}  

\begin{table}
\bibpunct{}{}{,}{n}{,}{,}
    \centering
    \caption{Solute uptake into the interfacial regions defined by the 10\%--90\% thresholds of the tetrahedral order parameter, $q(z)$. Uncertainties in the last digit are shown in parentheses. All concentrations are 1 M unless stated otherwise.}
\label{tab:uptake}
\begin{tabular}{ l | r }
                \midrule
Solute & interfacial uptake (M)\\
\midrule
DMSO & 0.17(6) \\
\ce{CH2(OH)2} & 0.27(8) \\
CHOOH &  0.23(7)\\
\ce{HCOONa} & 0.17(6) \\
0.5M \ce{HCOONa} & 0.07(4) \\
NaCl &  0.21(6)\\
NaF & 0.14(6)\\
\ce{NH4Cl} &  0.21(6)\\
0.5M \ce{NH4Cl} &  0.10(3)\\
\bottomrule
\end{tabular}
\end{table}

\subsection{Characterization of dynamics in the water environments}

\subsubsection{Hydrogen bond lifetimes}
Hydrogen bond jump correlation functions can be used to estimate the duration of hydrogen bonds in water.\cite{Laage:2006aa,Laage:2008aa} As hydrogen bond donor $i$ transitions between two hydrogen bond acceptors, $a$ and $b$, the associated occupancy variables transition simultaneously, $n_a: 1 \rightarrow 0 $ and $n_b: 0 \rightarrow 1$. The longevity of hydrogen bonds under different conditions can then be calculated via the jump time correlation function,
\begin{equation}
    C_\mathrm{jump}(t) = 1 - \langle n_{a}(0) n_{b}(t) \rangle
\end{equation}
where the angle brackets indicate averaging over all donors in the system (as well as averaging over initial times). Hydrogen bond jump dynamics can be further analyzed in terms of distance of the donor from a molecular inclusion $(r)$ or spatially to provide a measure of dynamic order near an interface,
\begin{eqnarray}
C_\mathrm{jump}(r,t) & = 1 - \langle n_{a}(0)n_{b}(t)\delta(r_{i}(0) - r) \rangle \\
C_\mathrm{jump}(z,t) & = 1 - \langle n_{a}(0)n_{b}(t)\delta(z_{i}(0) - z) \rangle
\end{eqnarray}

The resulting correlation functions can be fit with a triexponential decay function,\cite{Louden:2017aa}
\begin{equation}
C_\mathrm{jump}(t) \approx A_{s}e^{-t/\tau_{s}} + A_{m}e^{-t/\tau_{m}} + (1 - A_{s} - A_{m})e^{-t/\tau_{l}}.
\end{equation}
where $\{\tau_s, \tau_m , \tau_l\}$ are three timescales for decay and $\{A_s, A_m, (1-A_s-A_m)\}$ are the fractional contributions of each of these timescales to the overall decay.  The characteristic motions normally attributed to the three time scales are librational motion ($\tau_{s}$), angular jumps from breaking and reforming hydrogen  bonds with neighboring water molecules ($\tau_{m}$), and translational diffusion that breaks local water cages ($\tau_{l}$). 
Fitting the jump correlation function this way
provides significant insight into the contribution of various mechanisms to the decay of hydrogen bond lifetimes, and their time scales over a range of temperatures, solutes, and proximity to the ice surface. 
$C_\mathrm{jump}(\left\{r,z\right\}, t)$ may also be integrated to give characteristic decay times, providing a single time scale for hydrogen  bonding either as a function of distance from a molecular inclusion or as a function of spatial position relative to the ice interface.

Longer $\tau_\mathrm{jump}$ values correspond to more ice-like behavior, and have been utilized in previous work to arrive at a dynamic measure of the interfacial widths.\cite{Louden:2017aa}  Hydrogen bonds within the ice slab do not decay, so $\tau_\mathrm{jump} \rightarrow \infty$, but they change in the bulk liquid on the order of picoseconds. Spatially resolving $\tau_\mathrm{jump}$, therefore provides local information about the effects of molecular inclusions or the ice interface on local hydrogen bond dynamics,
\begin{equation}
\tau_\mathrm{jump}(\left\{r,z\right\}) = \int_0^\infty C_\mathrm{jump}(\left\{r,z\right\}, t) dt
\label{eq:intjump}
\end{equation}

\begin{figure}[]
    \centering
    \includegraphics[width=\linewidth]{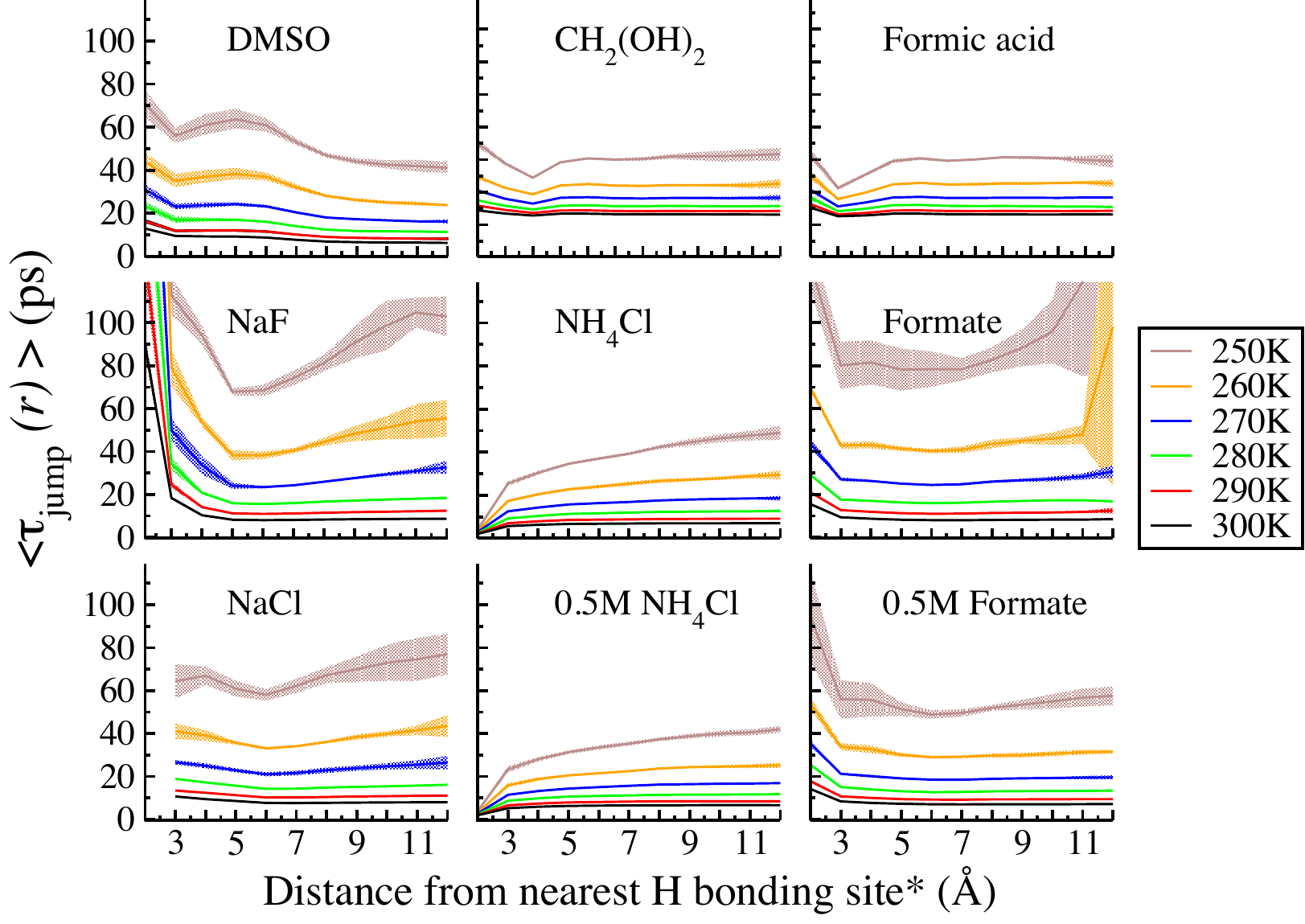}
    \caption{Integrated hydrogen bond jump times [Eq. \eqref{eq:intjump}] as a function of distance from the solute hydrogen bonding site (Cl- for NaCl). DMSO promotes a radially decreasing slowing effect in the surrounding water. \ce{NH4Cl} is the only ionic species in this study that does not slow the hydrogen bond dynamics near the molecular inclusion.}
    \label{fig:jumptimes(R)}
\end{figure}

Molecular inclusions that are net hydrogen bond \textit{acceptors} (and NaCl) slow the water dynamics more than inclusions that are net donors or molecules that are equitable participants in the hydrogen bonding network. The increase in hydrogen bonding lifetimes appears to correspond to the oversupply of acceptor sites, as 1M NaF and formate solutions exhibit the longest-lasting hydrogen bonds. For sodium formate, we observe a concentration dependence in the slowdown, which suggests cooperative effects between the solute molecules.

The presence of any of the solutes increases the hydrogen bond lifetime. The interfacial liquid follows the same trend as the bulk supercooled liquids -- solutes that are excess acceptors promote longer-lasting hydrogen bonds. We observe concentration dependence for both the formate and \ce{NH4Cl} systems, and the introduction of the ice surface disproportionately lengthens lifetimes in the DMSO and formate systems. Slower hydrogen bond dynamics appears to be a characteristic shared by the known cryoprotectant and deicing agents in both the interfacial and supercooled liquid systems. 

\begin{table}
\bibpunct{}{}{,}{n}{,}{,}
    \centering
    \caption{Hydrogen bond jump activation energies (in kcal mol$^{-1}$) for bulk supercooled solutions and supercooled interfacial liquids adjacent to ice.  All concentrations are 1 M unless otherwise stated, and all calculations were done using the TIP4P-Ice water model. Uncertainties in the final digit are shown in parentheses.}
\label{tab:EaSlope}
\begin{tabular}{ l | r r}
                \midrule
 
Solute & Bulk supercooled & Interfacial liquid\\
\midrule
Pristine & $5.2(2)$ & $8.6(2)$\\
DMSO & $5.7(2)$ & $8.0(6)$\\
\ce{CH2(OH)2} & $5.4(2)$ & $8.6(4)$ \\
CHOOH & $5.4(1)$ & $8.7(1)$ \\
\ce{HCOONa} & $6.7(2)$ & $8.1(6)$\\
0.5M \ce{HCOONa} & $6.0(2)$ & $8.5(3)$\\
NaCl & $6.2(2)$ & $8.4(4)$\\
NaF & $6.20(6)$ & $6.6(5)$\\
\ce{NH4Cl} & $5.2(1)$ & $8.6(3)$ \\
0.5M \ce{NH4Cl} & $5.2(1)$ & $8.44(8)$\\
\bottomrule
\end{tabular}
\end{table}

With these computed hydrogen bond jump times,  using an Arrhenius analysis, we calculated the activation energies associated with breaking and reforming hydrogen bonds.  While work by Piskulich \textit{et al.} found an activation energy under the extended jump model of 3.70 kcal/mol for SPC/E~\cite{Piskulich:2020aa} and 3.63 kcal/mol for TIP4P/2005~\cite{Piskulich:2021aa}, both at 298 K, TIP4P-Ice was not part of the tested models. Our pristine solution jump times are in agreement with the work of Baran \textit{et al.} on TIP4P-Ice at 273K.\cite{Baran:2023aa} The slower dynamics in TIP4P-Ice (and the significantly different melting point) result in higher jump time activation energies in this model compared to other water models. Generally, we find that the introduction of the ice sheet affects $E_a$ more than the identity of the solute, with the exception of the 1M NaF solution.

\subsection{Solid / liquid friction at ice--water interfaces}
\begin{figure}[]
    \centering
    \includegraphics[width=\linewidth]{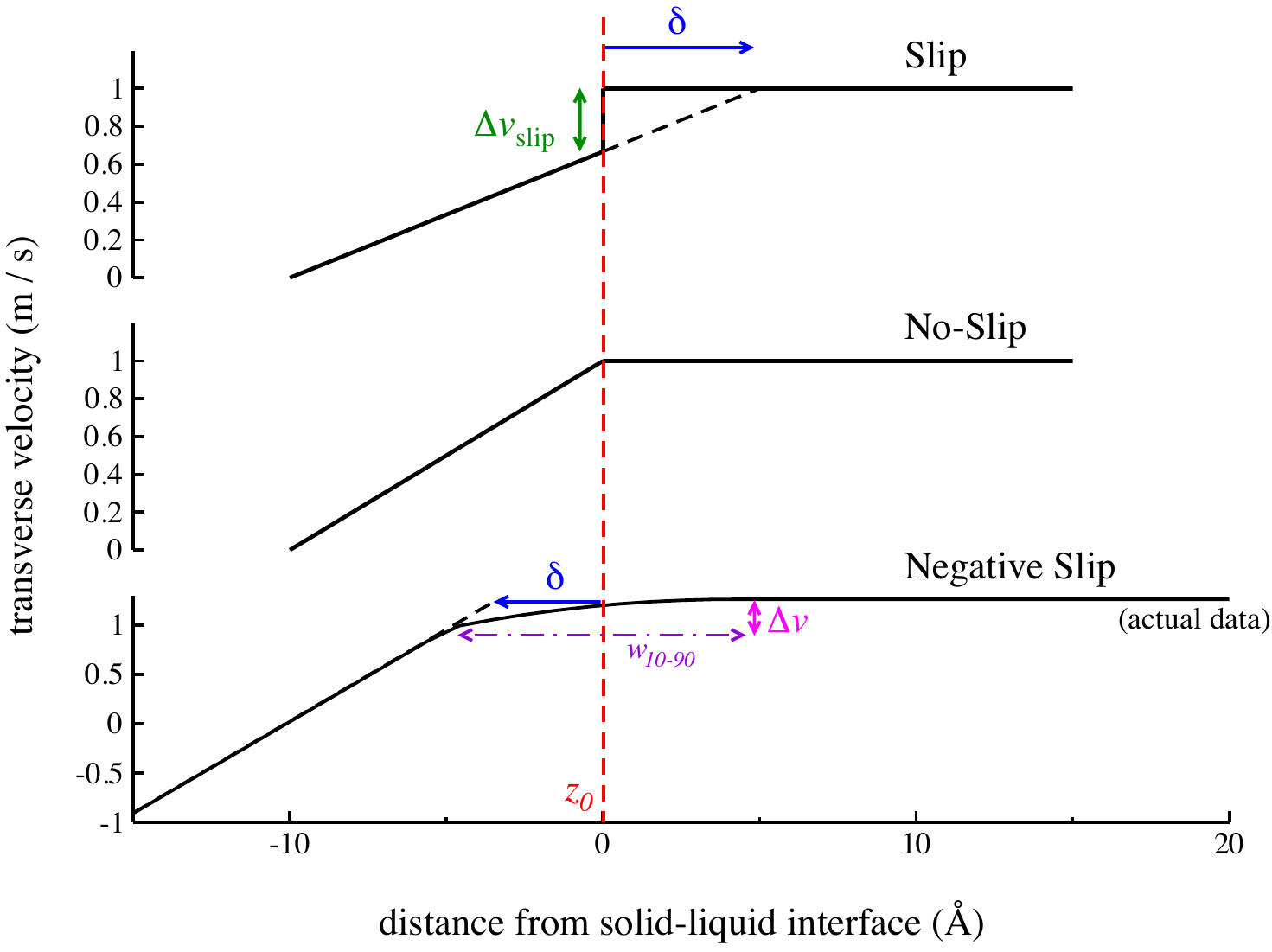}
    \caption{The different slip cases at solid/liquid interfaces. The velocity profile for the negative slip case uses averaged velocity data from Fig. \ref{fig:SoluteIceSim}}
    \label{fig:SlipLengthProfiles}
\end{figure}

Fluid flowing over a solid in no-stick boundary conditions is characterized by a slip length, $\delta$, projecting the tangential velocity of the liquid to zero at a location inside the solid (see Fig. \ref{fig:SlipLengthProfiles}).  For solid / liquid interfaces with weak interactions between the two phases, there is
very little drag imposed on the fluid, and the resulting interfacial liquid
velocity, $\Delta v_\mathrm{slip}$, can be significant. Under no-stick
boundary conditions, the extrapolated slip lengths can be quite large.  Balasubramanian and
Mundy related the slip length to an interfacial friction coefficient,
\begin{equation}
\kappa = \frac{\eta}{\delta}
\label{kappa1}
\end{equation}
where $\eta$ is the shear viscosity of the
liquid, and $\delta$ is the slip length.\cite{Balasubramanian:1999aa}

For solid / liquid interfaces where the liquid has \textit{strong} interactions with the solid, a larger frictional drag is imposed on the liquid near the interface. As a result, the slip lengths become smaller, and systems can reach ``no-slip'' boundary conditions where the projected tangential velocity goes to zero at the interface. The slip length approaches zero in these cases, so no-slip boundaries pose a problem for
Eq. \eqref{kappa1}, as $\kappa$ asymptotically goes to infinity as
$\delta \rightarrow 0$.  

In the interfaces studied here (ice shearing through liquid water), the 
solid / liquid interactions are so strong that the tangential 
velocities vanish \textit{before} reaching the solid / liquid boundary,
and some of the liquid is pulled along with the solid.
Because the tangential velocity profile of the liquid extrapolates to 
the wall velocity several molecular layers before reaching the solid, 
these systems exhibit a \textit{negative} slip length, and the solid / liquid friction coefficient defined in Eq. \eqref{kappa1} becomes meaningless. 

The solid / liquid friction coefficient may also be defined
using the velocity drop across the interface, rather than the length
scale over which this drop occurs. We can relate the imposed shear
stress to the relative tangential velocity of the fluid in the
interfacial region,\cite{Louden:2017aa}
\begin{equation}\label{Shenyu-13}
j_{z}(p_{x}) = \kappa \Delta v
\end{equation}
where $\Delta v = v_{x}(\mathrm{solid}) - v_{x}(\mathrm{liquid})$ is
the difference in transverse velocity between points that are
unambiguously on the solid and liquid sides of the interface, and $j_z(p_x)$ is the imposed momentum flux. In slip
boundary conditions, the two definitions of the solid--liquid friction coefficient are identical, but
Eq. \eqref{Shenyu-13} provides a direct analogy to
expressions for the interfacial thermal conductance $(G)$,
\begin{equation}
J_q = G~ \Delta T,
\end{equation}
where $J_q$ is a thermal flux and the temperature drop is measured
across an interface of finite width. By analogy, $\kappa$ is
a transport coefficient that measures \textit{interfacial momentum
  conductance} across an interface of finite width.

To estimate $\kappa$, tangential velocity profiles from each simulation were fit using a
piecewise function that is both continuous and continuously
differentiable (see the supplementary material). To arrive at estimates of
the interfacial velocities, these fits were queried at locations on
either side of the Gibbs dividing surface defining the solid/liquid interface,
\begin{align}
v_{x}(\mathrm{solid}) & = v_{x} \left( z_0 + \frac{w_\mathrm{10-90}}{2}\right)  \label{eq:vx1}\\
v_{x}(\mathrm{liquid}) & = v_{x} \left( z_0 - \frac{w_\mathrm{10-90}}{2}\right). \label{eq:vx2}
\end{align}
Here, $z_0$ is the location of the Gibbs dividing surface and $w_\mathrm{10-90}$ is the width of the interface. Both these parameters must be determined by a structural order parameter.

In this work, the locations of the Gibbs dividing surfaces and the interfacial widths were obtained using fits to the spatial profile of tetrahedrality.  The fit equations have been adjusted to account for the imposed thermal gradient,
\begin{equation}
    q(z) = q_\mathrm{liq} + \frac{q_\mathrm{ice} - q_\mathrm{liq}}{2} \left( \tanh\left({\frac{z - l}{w}}\right) - \tanh\left({\frac{z-r}{w}}\right) \right) + \beta \left| z - \frac{r+l}{2} \right|
\end{equation}
Here, $q_\mathrm{ice}$ and $q_\mathrm{liquid}$ are the bulk behavior of tetrahedrality in the two phases, while $l$ and $r$ are the locations of the left and right Gibbs dividing surfaces, respectively.  $w$ is the structural width of this interface, and $\beta$ is a parameter that accounts for the effects of the small thermal gradients that are imposed during shearing simulations to prevent shear-induced melting.

When fitting velocity profiles with 1\AA\ resolution, some bins inside the ice had very low populations and significant noise.  We fit velocity profiles only for regions that had densities $\rho(z) > 0.25 \mathrm{~g~cm}^{-3}$. Viscosities of the interfacial liquid, $\eta$, were also calculated from the imposed momentum flux, $j_z(p_x)$, and the slope of the velocity profile in the liquid portion of the simulation,
\begin{equation}
    j_z(p_z) = -\eta \left(\frac{d v_x}{d z} \right).
\end{equation}
Sample velocity and tetrahedrality profiles can be seen in Fig.  \ref{fig:SoluteIceSim}. The calculated interfacial friction coefficients $(\kappa)$ and solution-phase viscosities $(\eta)$ for each solute system are given in Table \ref{tab:eta and KappaVal}. Baran \textit{et al.} reported that the $\eta$ of pure TIP4P-Ice is 3.585 mPa s at 273K.\cite{Baran:2023aa} At the lower temperature of our bulk pristine system (269.7K) we expect slightly larger $\eta$.

\begin{table}[]
    \centering
    \caption{Solution-phase shear viscosities ($\eta$) and solid--liquid friction coefficients ($\kappa$) for solutions in contact with ice-Ih basal facets. All concentrations are 1 M unless otherwise stated. Uncertainties in the last digit are given in parentheses.}
\label{tab:eta and KappaVal}
\begin{tabular}{ l | c | c | c | c | c}
                \toprule
\multirow{2}*{Solute} & \multicolumn{2}{|c}{Bulk solution} & \multicolumn{3}{|c}{Interfacial liquid} \\   \cmidrule{2-3} \cmidrule{4-6}
 & T (K) & $\eta~\mathrm{(mPa \cdot s)}$ & T (K) & $\eta~\mathrm{(mPa \cdot s)}$ & $\kappa~ \mathrm{(amu~\angstrom^{-2}~fs^{-1})}$  \\
\midrule
Pristine & $269.7(2)$ & $ 3.9(1)$ & $271.3(4)$ & $ 6.2(6)$ & $ 0.38(7)$ \\
DMSO & $270.0(2)$ & $ 5.8(2) $ & $272.4(6)$ & $11.8(5)$ & $ 0.9(3)$ \\
\ce{CH2(OH)2} & $269.6(3)$ & $ 4.4(1)$ & $272.0(5)$ &  $ 8.6(8)$ & $ 0.7(3)$ \\
CHOOH &  $269.8(3)$ & $4.3(2)$ & $271.7(2)$ &  $ 7(1)$ & $ 0.4(1)$ \\
\ce{HCOONa} &  $270.0(3)$  & $8.3(3)$ & $272.5(3)$ &  $12.2(6)$ & $0.7(2)$ \\
0.5M \ce{HCOONa} & $269.9(4)$ & $5.7(2)$ & $272.1(7)$ &  $8.7(6)$ & $0.4(1)$ \\
NaCl & $269.6(6)$ & $7.0(3)$ & $271.7(5)$ &  $11(1)$ & $0.33(3)$ \\
NaF & $270.2(1)$ & $9.8(3)$ & $272.6(4)$ &  $16(4)$ & $ 0.4(2)$ \\
\ce{NH4Cl} & $269.9(4)$ & $4.3(1)$ & $271.7(6)$ &  $7.9(5)$ & $ 0.4(1)$ \\
0.5M \ce{NH4Cl} & $270.0(3)$ & $4.3(1)$ & $270.9(1)$ &  $6.9(3)$ & $ 0.5(1)$ \\
\bottomrule
\end{tabular}
\end{table}

Reported temperatures in the interfacial systems are averages only of the liquid-phase portion of the systems. These temperatures are higher in the interfacial systems primarily due to the imposed thermal gradient which was used to keep the ice below 270 K. The addition of ice crystals and solute increases $\eta$ in relation to the pristine systems despite the higher temperatures of the interfacial liquid. As seen in the hydrogen bond jump lifetimes, excess hydrogen bond acceptors and NaCl have the largest impact on the viscosity of the liquid. Only DMSO and 1M formate have statistically significant different friction values $(\kappa)$ relative to the pristine system.

\begin{figure}[]
    \centering
    \includegraphics[width=\linewidth]{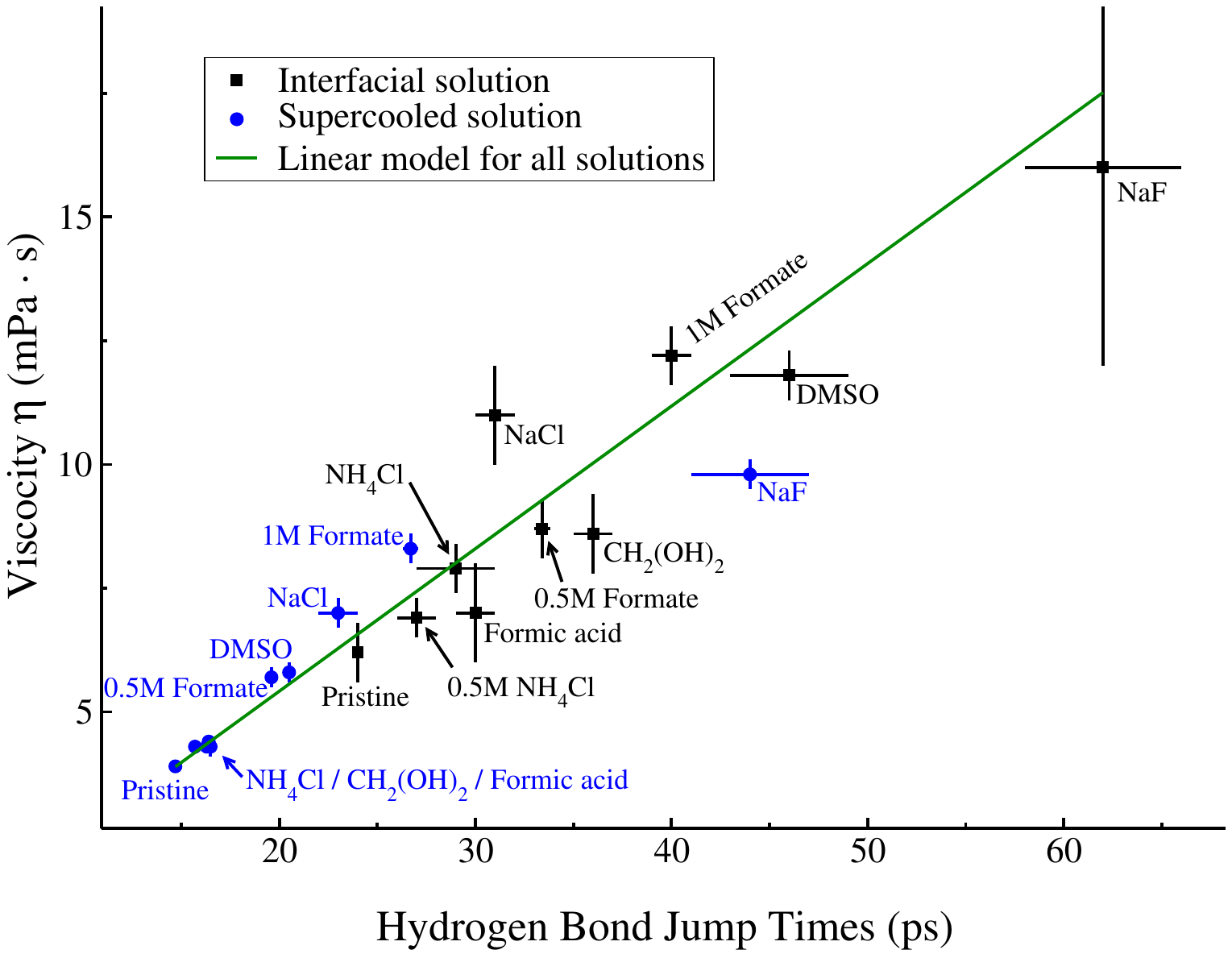}
    \caption{Correlation between shear viscosity ($\eta$) and hydrogen bond jump times for both supercooled and interfacial solutions at temperatures listed in Table 5.  All concentrations are 1 M unless stated.}
    \label{fig:EtavJumpTime}
\end{figure}

We observe a linear correlation between the hydrogen bond jump time measured in a liquid and the shear viscosity of the same liquid (see Fig. \ref{fig:EtavJumpTime}).
It is not surprising that the time required for water hydrogen bonds to switch partners is strongly correlated with the viscosity in the liquid, as diffusion and viscosity are generally inversely proportional to each other.  We note that the DMSO and \ce{CH2(OH)2} solutions appear to be impacted most by the presence of the ice crystal.

\section{Conclusion}
Utilizing molecular dynamics, we have interrogated the structural and dynamic solution properties of a cryoprotectant (DMSO), a polyalcohol [\ce{CH2(OH)2}], a carboxylic acid in protonated and deprotonated forms (\ce{HCOOH} and \ce{HCOONa}), an ammonium salt (\ce{NH4Cl}), and two alkyl halide salts (\ce{NaCl} and \ce{NaF}) in bulk solution and in proximity to ice. In addition to the different impacts on ice formation, these solutes represent a range of hydrogen-bonding behaviors, oxidation states, and chemical analogies to amino acid groups. Structural order parameters and time correlation functions show a significant impact on the water hydrogen bonding networks, at least in the TIP4P-Ice water model that was used in this study. At short distances (within 4 \AA) all of the solutes induce structural disorder, but when considering the entire solution, the deicing agents and DMSO increase the orientational ordering and increase hydrogen bond lifetimes for water molecules found 0.6--1 nm from solute hydrogen bonding sites (\ce{Cl-} for NaCl). Notably, the ice nucleation enhancer, \ce{NH4Cl}, decreased structural order and hydrogen bond lifetimes for water molecules found at these distances. We find a correlation between net water hydrogen bond statistics (donor / acceptor) and the orientational order parameter, where the hydrogen bond acceptors (NaF, DMSO, and HCOONa) increase structural order and the hydrogen bond donor (\ce{NH4Cl}) decreases structural order.
    
When ice is introduced into these solutions, the behavior of water is significantly altered. Generally, the presence of the nearby ice interface enhances structural order, lengthens hydrogen bond lifetimes, and increases shear viscosity relative to the supercooled liquid solutions at the same temperatures. The ionic compounds caused a decrease in order relative to the molecular solutes. Solutions with hydrogen bond acceptors exhibit significantly longer-lived hydrogen bonds, while solutions with donors have shorter lifetimes. That said, the presence of any solute lengthens hydrogen bond lifetimes relative to the pristine interface.  When looking at shear viscosities, we observed a nearly linear correlation with hydrogen bond jump times across all solutions studied.

Using tetrahedrality as the spatial order parameter distinguishing solid from liquid, ionic solutes were found to increase the interfacial widths by $\sim 2$ \AA\ relative to the pristine and molecular systems. However, interfacial widths measured using translational or dynamic order parameters did not have this clear delineation.
    
The deicers exhibit relatively high activation barriers $(E_a)$ for breaking hydrogen bonds. However, in the presence of the ice interface,  the activation barriers converge to similar values (with the exception of \ce{NaF}).  We find that the presence of the ice surface has a larger impact on H-bond activation than the solute identity.  Only DMSO and \ce{HCOONa} exhibited statistically significant increases in interfacial friction coefficients. This suggests an interesting role for formate salts as potential runway and roadway deicers. 

One interesting question is whether cooperative effects and molecular crowding have an impact on solutions with multiple solutes. The closest our study can come to answering this question is via the two solutes simulated at multiple concentrations. We note that there are significant differences in tetrahedral ordering for the 0.5M and 1M \ce{NH4Cl} solutions, suggesting that multiple nearby donors may reduce local ordering in water. Similarly, the hydrogen bonding jump times are significantly higher for the 1M \ce{HCOONa} solution relative to the 0.5M solution, suggesting that multiple nearby anionic acceptors may have an increased impact on the dynamics of water.

Our primary findings are as follows: (1) the deicing agents studied here exhibit slower dynamics and act to disorder the water networks in the presence of ice, (2) DMSO, a known cryoprotectant, causes slower hydrogen bonding but acts to structurally order the surrounding water, and (3) the nucleation promoter \ce{NH4Cl} creates relatively rapid hydrogen bonding dynamics in a disordered water network. This study reinforces that a combination of characteristics, notably ionic character and a hydrogen bond acceptor--donor imbalance, works collectively to aid in deicing, cryopreservation, or promotion of ice nucleation.  We also emphasize that the presence of the ice interface affects the relative importance of these characteristics.

\section{Supplementary material}
The supplementary material contains force field information, anion vs cation uptake data, temperature dependence of tetrahedral and translational order parameters, hydrogen bond jump times, and their respective interfacial widths.

\section{Acknowledgments}
 Support for this project was provided by the National Science Foundation under grant CHE-1954648. Computational time was provided by the Center for Research Computing (CRC) at the University of Notre Dame. The authors also recognize contributions to this project by Patrick Louden.

\section{Author Declarations}

\subsection{Conflict of Interest}
The authors have no conflicts to disclose. 

\subsection{Author Contributions}
This work was made available through contributions from all authors. B.M.H. and J.D.G. conceived and designed the simulations. B.M.H. and J.K.S. performed the simulations. B.M.H and J.D.G. analyzed the data. B.M.H. and J.D.G. wrote and edited the manuscript. J.D.G. secured financial support for the research.

\section{Data Availability}
The data that support the findings of this study are available within the article and its supplementary material. All simulations utilized the OpenMD molecular dynamics engine (see \href{https://github.com/OpenMD/OpenMD}{github.com/OpenMD/OpenMD}),\cite{Drisko:2024aa} which is available under a BSD 3-clause license. All analysis tools utilized here have been built into OpenMD. 

\bibliography{main}
\end{document}


\clearpage
\pagebreak
\setcounter{page}{1}
\onecolumngrid
\widetext
\begin{center}
    \textbf{{\large Supplemental Material for: ``Molecular Dynamics of Ice-Active Solutions at Ice-Water Interfaces''}}\\[1em]
Benjamin M. Harless, Jasmine K. Sindelar, and J. Daniel Gezelter \\
251 Nieuwland Science Hall, Department of Chemistry and Biochemistry\\
University of Notre Dame, Notre Dame, Indiana 46556
\end{center}

\section{Forcefield parameters}
Force field parameters for the \ce{CH2(OH)2}, \ce{CHOOH}, and \ce{HCOONa} anion species were adapted from the Generalized Amber Force Field (GAFF)\cite{Wang:2004aa} with charges on the \ce{CH2(OH)2} adapted from the work of Fennell, Wymer, and Mobley, which provides accurate descriptions of solvation free energies for alcohols.\cite{Fennell:2014aa}  DMSO was modeled using the  CHARMM General Force Field (CGenFF) parameters provided by Strader and Feller\cite{Strader:2002aa} which has been previously shown to give accurate IR spectra when solvated in water.\cite{Lee:2022aa}  The ammonium cation parameters were adapted from the work of Kashefolgheta and Vila Verde, \cite{Kashefolgheta:2017aa} while the monatomic ions were modeled using the general 12-6 Lennard-Jones parameters from Li, Song, and Merz.\cite{Li:2015aa}
Rigid TIP4P-Ice water molecules\cite{Abascal:2005aa} were used as models for the solid ice and the liquid water phase for this work.
The damped shifted force (DSF) kernel by Fennell and Gezelter was used for all long-range electrostatic interactions with a damping parameter ($\alpha$) of 0.18 \AA{}$^{-1}$.\cite{Fennell:2006aa}   Table \ref{tab:atypes} provides the mass, charge, and non-bonded (Lennard-Jones) parameters for the species used in this work. 

\renewcommand{\arraystretch}{0.75} 
\begin{table}
\bibpunct{}{}{,}{n}{,}{,}
    \centering
    \caption{Atomic Parameters}
\label{tab:atypes} 
\small
\begin{tabular}{ r | r r r r r}
                \midrule
AtomType & Mass (amu) & Charge ($e^-$) & $\epsilon$ (kcal/mol) & $\sigma$ (A) & Source \\
\midrule 
\multicolumn{5}{l}{DMSO} & Ref. \cite{Strader:2002aa} \\ \midrule
C & 12.01 & -0.148 & 0.078 & 3.634867 & \\
S & 32.06 & 0.312 & 0.350 & 3.563595 & \\
O & 16.0 & -0.556 & 0.120 & 3.029056 & \\
H & 1.008 & 0.090 & 0.024 & 2.387609 & \\ \midrule
\multicolumn{5}{l}{Formate anion} & Ref. \cite{Wang:2004aa} \\ \midrule
C & 12.01 & 0.8489 & 0.0988 & 3.3152\\
HC & 1.008 & -0.1989 & 0.0161 & 2.4473\\
O2 & 16.0 & -0.8250 & 0.1463 & 3.0481\\ \midrule
\multicolumn{5}{l}{\ce{CH2(OH)2}} & Refs. \cite{Wang:2004aa} and \cite{Fennell:2014aa} \\ \midrule
C & 12.01 & 0.453622 & 0.109 & 3.39967\\
HC & 1.008 & 0.071361 & 0.0157 & 2.29317\\
OH & 16.0 & -0.782810 & 0.2021 & 3.21990\\
HO & 1.008 & 0.484638 & 0.0 & 0.0\\ \midrule
\multicolumn{5}{l}{Formic Acid} & Ref. \cite{Wang:2004aa}\\ \midrule
C1 & 12.01 & 0.7119 & 0.0988 & 3.3152\\
HO & 1.008 & 0.4721 & 0.0047 & 0.5379\\
OH & 16.0 & -0.6509 & 0.0930 & 3.2429\\
O2 & 16.0 & -0.5786 & 0.1463 & 3.0481\\
HC & 1.008 & 0.0455 & 0.0161 & 2.4473\\ \midrule
\multicolumn{5}{l}{Ammonium cation} & Ref. \cite{Kashefolgheta:2017aa} \\ \midrule
N  & 14.01 & -0.706858 &  0.17  &  3.25  & \\
H  & 1.008 & 0.426714 & 0.0157  & 1.06908 & \\ \midrule
\multicolumn{5}{l}{Ionic Species} & Ref. \cite{Li:2015aa} \\ \midrule
\ce{Na+} & 22.990 & 1.0 & 0.02909167 & 2.610 & \\
\ce{Cl-} & 35.453 & -1.0 & 0.16573832 & 3.099 & \\
\ce{F-} & 18.998403 & -1.0 & 0.5315665 & 3.852 &\\ \midrule
\multicolumn{5}{l}{TIP4P-Ice} & Ref. \cite{Abascal:2005aa} \\ \midrule
O & 16.0 & 0.0 & 0.210842 & 3.1668 & \\
H & 1.008 & 0.5897 & 0.0 & 0.0 & \\
EP & 0.0 & -1.1794 & 0.0 & 0.0 & \\
\bottomrule
\end{tabular}
\end{table}

In Table \ref{tab:bond}, harmonic bonds are described by
\begin{equation}
    V_\text{bond}(r) = \frac{k_b}{2} (r - b_0)^2
\end{equation}
where $k_b$ is the force constant and $b_0$ is the equilibrium bond length. Water is simulated as a rigid body so no harmonic bond parameters are needed. The base atom types are shown here when applicable, so C refers to all carbon atoms that are not accounted for in the other listed harmonic bond parameters (e.g., C1) and O refers to all oxygen atoms that are not otherwise listed (O2 and OH). 

\begin{table}
\bibpunct{}{}{,}{n}{,}{,}
    \centering
    \caption{Harmonic Bond Parameters}
\label{tab:bond}
\begin{tabular}{ c c | c c c}
                \midrule
Atom1 & Atom2 & $b_0$ (\AA) & $k_b$ (kcal/mol) & Source \\
\midrule
\multicolumn{4}{l}{DMSO} & Ref. \cite{Strader:2002aa}\\ \midrule
C & S & 1.80 & 240 & \\
C & HC & 1.11 & 322 & \\
O & S & 1.53 & 540 & \\
\midrule
\multicolumn{4}{l}{Formate anion} & Ref. \cite{Wang:2004aa} \\ \midrule
C & O2 & 1.218 & 652.6 & \\
C & HC & 1.105 & 361.8 & \\
\midrule
\multicolumn{4}{l}{\ce{CH2(OH)2}} & Ref. \cite{Wang:2004aa} \\ \midrule
HC & C & 1.096 & 377.3 & \\
C & OH & 1.423 & 293.4 & \\
OH & HO & 0.973 & 563.5 & \\
\midrule
\multicolumn{4}{l}{Formic Acid} & Ref. \cite{Wang:2004aa} \\ \midrule
C1 & HC & 1.105 & 361.8 &\\ 
OH & HO & 0.973 & 563.5 & \\
C1 & OH & 1.341 & 383.1 &\\ 
C1 & O2 & 1.218 & 652.6 &\\
\midrule
\multicolumn{4}{l}{Ammonium cation} & Ref. \cite{Kashefolgheta:2017aa}\\ \midrule
N & H & 1.033 & 738.0019 &\\
\bottomrule
\end{tabular}
\end{table}

In Table \ref{tab:bend}, harmonic bends are in the form
\begin{equation}
V_{\text{bend}}(\theta) = \frac{k_\theta}{2}(\theta - \theta_0)^2
\end{equation}
where $\theta$ describes the angle between bonded atoms of type $i-j-k$, where $j$ is the central atom.

\begin{table}
\bibpunct{}{}{,}{n}{,}{,}
    \centering
    \caption{Harmonic Bend Parameters}
\label{tab:bend}
\begin{tabular}{ c c c | c c c }
                \midrule
 
Atom1 & Atom2 & Atom3 & $\theta_0~(^\circ)$ & $k_\theta \mathrm{~(kcal~mol^{-1}~rad^{-2})} $ & Source\\
\midrule
\multicolumn{5}{l}{DMSO}  & Ref. \cite{Strader:2002aa}\\ \midrule
HC & C & HC & 108.4 & 35.5 &\\
HC & C & S & 111.3 & 46.1 &\\
C & S & O & 106.75 & 79.0 &\\
C & S & C & 95.0 & 34.0 &\\
\midrule
\multicolumn{5}{l}{Formate anion} & Ref. \cite{Wang:2004aa} \\ \midrule
O2 & C & O2 & 130.25 & 118.8&\\
HC & C & O2 & 123.65 & 65.9&\\
\midrule
\multicolumn{5}{l}{\ce{CH2(OH)2}} & Ref. \cite{Wang:2004aa}\\ \midrule
C & OH & HO & 107.26 & 49.0&\\
OH & C & OH & 109.90 & 110.7&\\
OH & C & HC & 109.43 & 62.8&\\ 
HC & C & HC & 110.2 & 38.5&\\
\midrule
\multicolumn{5}{l}{Formic Acid} & Ref. \cite{Wang:2004aa}\\ \midrule
C1 & OH & HO & 106.57 & 51.6&\\
OH & C1 & O2 & 122.01 & 115.7&\\
OH & C1 & HC & 109.49 & 65.3&\\
O2 & C1 & HC & 123.64 & 65.9&\\
\midrule
\multicolumn{5}{l}{Ammonium cation} & Ref. \cite{Kashefolgheta:2017aa}\\ \midrule
H & N & H & 109.5 & 81.0397 & \\
\bottomrule
\end{tabular}
\end{table}

In Table \ref{tab:torsions}, CHARMM-style torsions take the form 
\begin{equation}
V_{\text{torsion}}(\phi) = \sum_n K_n (1 + \cos(n\phi - \delta_n))
\end{equation}
where $K_n$ is the amplitude and $\delta_n$ is the phase angle corresponding to a cosine contribution with periodicity $n$. 

\begin{table}
\bibpunct{}{}{,}{n}{,}{,}
    \centering
    \caption{Torsion parameters}
\label{tab:torsions}
\begin{tabular}{ c c c c | c c c c}
                \midrule
 
Atom1 & Atom2 & Atom3 & Atom4 & $K_n \mathrm{(kcal~mol^{-1})}$ & $n$ & $\delta (^\circ)$ & Source \\
\midrule
\multicolumn{7}{l}{DMSO} & Ref. \cite{Strader:2002aa}\\ \midrule
HC & C & S & O & 0.2 & 3 & 0 &\\
HC & C & S & C & 0.2 & 3 & 0 &\\
\midrule
\multicolumn{7}{l}{\ce{CH2(OH)2}} & Ref. \cite{Wang:2004aa}\\ \midrule
OH & C & OH & HO & 1.57 & 2 & 0 &\\
HC & C & OH & HO & 0.167 & 3 & 0 &\\
\midrule
\multicolumn{7}{l}{Formic Acid} & Ref. \cite{Wang:2004aa}\\ \midrule
HC & C1 & OH & HO & 2.3 & 2 & 180 &\\
O2 & C1 & OH & HO & 1.9 & 1 & 0 &\\
O2 & C1 & OH & HO & 2.3 & 2 & 180&\\
\bottomrule
\end{tabular}
\end{table}

For inversions centered on $sp^2$-hybridized atoms with three satellites, the improper cosine torsion form is used for the formate and formic acid ion moieties,
\begin{equation}
V_{\text{inversion}}(\omega) = \sum_n K_n \left( 1 + cos(n
  \omega - \delta_n) \right),
\end{equation}
where $\omega$ is an improper torsion angle described with the central atom in position 3 of a standard torsion, and the satellite atoms in positions 1, 2, and 4.

\begin{table}[]
\bibpunct{}{}{,}{n}{,}{,}
    \centering
    \caption{Inversion Parameters}
\label{tab:inversion}
\begin{tabular}{ r | l l l l | c c c | c}
                \midrule
 
Molecule & Central Atom & Atom2 & Atom3 & Atom4 & $K_n$ & $n$ & $\delta (^\circ)$ & Source\\
\midrule
Formate & C & HC & O2 & O2 & 2.2 & 2 & 180 & Ref. \cite{Wang:2004aa}\\
 
Formic Acid & C1 & OH & O2 & HC & 2.2 & 2 & 180 & Ref. \cite{Wang:2004aa} \\
\bottomrule
\end{tabular}
\end{table}

\clearpage

\section{System Construction}

A bulk ice crystal ($44.9 \angstrom \times 46.68 \angstrom \times 44.04 \angstrom$) was also simulated at the same temperatures.  This crystal structure was generated using an ice-Ih unit cell proposed by Hirsch and Ojam\"{a}e (structure 6).\cite{Hirsch:2004aa}  All simulations of ice structures have proton ordering on the length scale of the periodic box, but the crystal used here has proton translational ordering on a smaller length scale than other ice studies. 

\section{Temperature dependence of Hydrogen Bond Statistics}

\begin{table}[h]
\bibpunct{}{}{,}{n}{,}{,}
    \centering
    \caption{Difference in Bulk water hydrogen bonding $(\langle N^{\ce{H2O}}_\mathrm{acceptor} \rangle - \langle N^{\ce{H2O}}_\mathrm{donor} \rangle)$ as a function of temperature. For a single solute, there is no systematic temperature dependence.}
\label{tab:TempHbond}
\begin{tabular}{ l | c c c c c c c}
                \midrule
 
Solution & 250K & 260K & 270K & 280K & 290K & 300K \\
\midrule
1M DMSO & -0.046 & -0.046 & -0.046  & -0.046 & -0.046  & -0.046  \\
1M \ce{CH2(OH)2} & -0.017 & -0.017  & -0.017 & -0.017  & -0.017  & -0.017 \\
1M CHOOH & -0.011  & -0.012 & -0.012  & -0.013  & -0.013  & -0.014 \\
1M \ce{HCOONa} & -0.117  & -0.116  & -0.115  & -0.117  & -0.116  & -0.114 \\
0.5M \ce{HCOONa} & -0.053  & -0.054  & -0.055  & -0.055  & -0.056  &  -0.057\\
1M NaF  & -0.107 & -0.107 & -0.111  & -0.110  & -0.112  & -0.112 \\
1M \ce{NH4Cl} & 0.061  & 0.060  & 0.060  & 0.059  & 0.059  & 0.058 \\
0.5M \ce{NH4Cl} & 0.030 & 0.030  & 0.030  & 0.030  & 0.030 & 0.030 \\
\bottomrule
\end{tabular}
\end{table}

\clearpage

\section{Temperature Dependence of Tetrahedral Ordering}

The following figures include the average bulk tetrahedral order parameters, $\langle q \rangle$, computed across a range of temperatures tested for both the supercooled liquid and interfacial systems.

\begin{figure}[h]
    \centering
    \includegraphics[width=\linewidth]{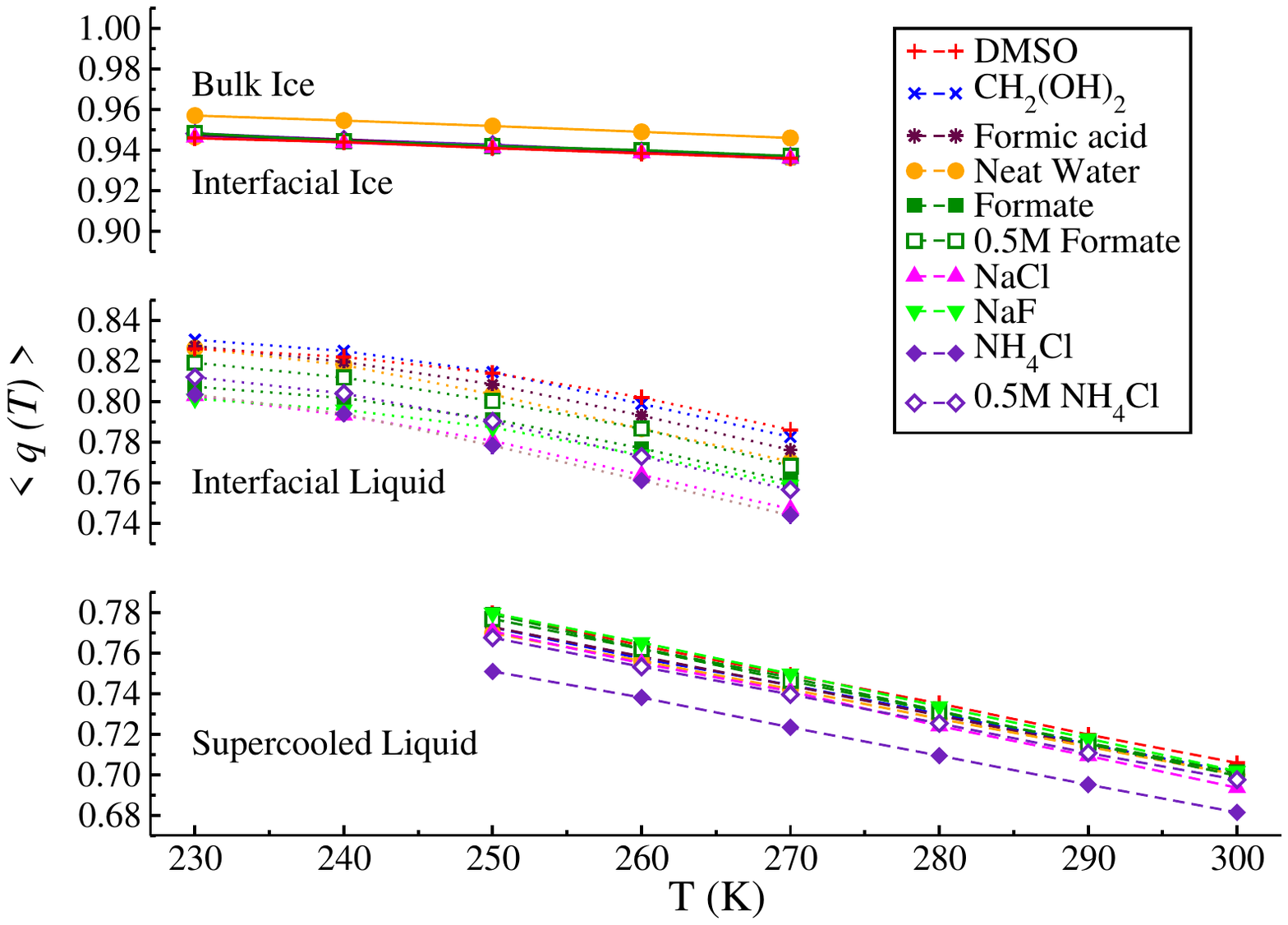}
    \caption{Mean tetrahedral order parameters, $\langle q \rangle$, as a function of temperature for the bulk supercooled solutions and interfacial ice / water systems. \ce{NH4Cl} is the most significant `structure breaking' solute, matched only by NaCl in the interfacial liquid. The confining effect of the ice surface alters the solution phase tetrahedrality relative to bulk supercooled liquids at the same temperatures.}
    \label{fig:average tet data}
\end{figure}  

\begin{table}[h]
\bibpunct{}{}{,}{n}{,}{,}
    \centering
    \caption{Interfacial widths (in \AA) calculated using the tetrahedal order parameter. We did not observe temperature dependence of these widths so an average over all temperatures was used in figure 5 of the main paper.}
\label{tab:TempQiW1090}
\begin{tabular}{ l | c c c c c c }
                \midrule
 
Solution & 230K & 240K & 250K & 260K & 270K\\
\midrule
Pristine & $8(1)$ & $8.6(6)$ & $10(1)$ & $9.2(4)$ & $8.6(3)$\\
1M DMSO & $8.8(7)$ & $9.9(6)$ & $10(1)$ & $8.2(6)$ & $9.7(6)$\\
1M \ce{CH2(OH)2} & $9(1)$ & $9.1(7)$ & $10(1)$ & $10(1)$ & $9.4(6)$\\
1M CHOOH & $9.7(3)$ & $9.4(2)$ & $10.1(9)$ & $10(1)$ & $9.6(8)$\\
1M \ce{HCOONa} & $11.0(7)$ & $10.6(5)$ & $11.2(6)$ & $11.3(7)$ & $11(1)$\\
0.5M \ce{HCOONa} & $10(1)$ & $10.5(9)$ & $11(2)$ & $10.5(7)$ & $ 10(1)$\\
1M NaCl & $11(2)$ & $12(1)$ & $11(1)$ & $12(1)$ & $11(1)$\\
1M NaF  & $11(2)$ & $10(2)$ & $11(2)$ & $11(2)$ & $11(2)$\\
1M \ce{NH4Cl} & $12(2)$ & $11(2)$ & $11(2)$ & $11(2)$ & $11(2)$\\
0.5M \ce{NH4Cl} & $11(2)$ & $11(1)$ & $11(1)$ & $10.6(8)$ & $11(1)$\\
\bottomrule
\end{tabular}
\end{table}

\clearpage

\section{Temperature Dependence of Translational Ordering}

The following figures include the average bulk translational order parameters, $\langle t \rangle$, calculated across the range of temperatures for both the supercooled liquid and interfacial systems.

\begin{figure}[h]
    \centering
    \includegraphics[width=\linewidth]{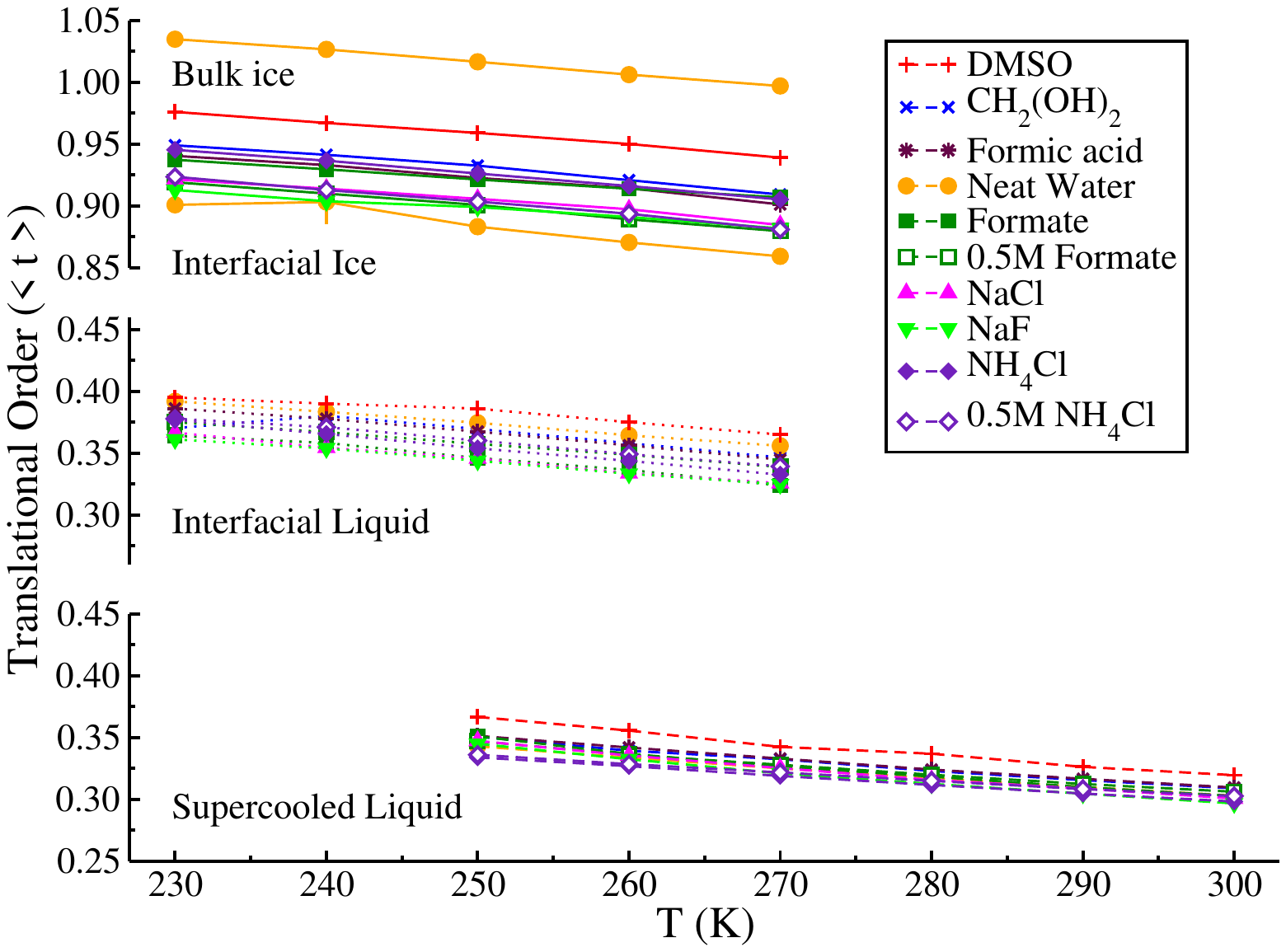}
    \caption{Average Translational order parameters as a function of temperature. DMSO is notable for both exhibiting the largest impact and increasing translational ordering in the solution relative to the pristine sample in the interfacial liquid. All solutes increase the translational order in the interfacial ice relative to the pristine system.}
    \label{fig:Translational Order solution data}
\end{figure}

\begin{table}[h]
\bibpunct{}{}{,}{n}{,}{,}
    \centering
    \caption{Interfacial widths (in \AA) calculated using the translational order parameter.  We did not observe strong temperature dependence, so an average over all temperatures was used in figure 5 of the main paper.}
\label{tab:TempTkW1090}
\begin{tabular}{ l | c c c c c c c}
                \midrule
 
Solution & 230K & 240K & 250K & 260K & 270K \\
\midrule
Pristine & $4.7(4)$ & $5.7(7)$ & $5.9(3)$ & $5.7(2)$ & $5.4(2)$\\
1M DMSO & $6.2(4)$ & $6.3(3)$ & $6.5(4)$ & $6.4(6)$ & $6.7(8)$\\
1M \ce{CH2(OH)2} & $6.2(2)$ & $6.1(2)$ & $7.1(5)$ & $6.8(7)$ & $6.8(5)$\\
1M CHOOH & $5.4(5)$ & $6.4(2)$ & $6.5(3)$ & $6.5(8)$ & $6.5(8)$\\
1M \ce{HCOONa} & $8(2)$ & $7.1(6)$ & $7.9(8)$ & $8(1)$ & $8(1)$ \\
0.5M \ce{HCOONa} & $ 5.8(5)$ & $6.1(9)$ & $6.7(7)$ & $6.8(3)$ & $6.3(6)$\\
1M NaCl & $6.2(6)$ & $7.4(6)$ & $7.3(5)$ & $7.6(5)$ & $6.7(5)$ \\
1M NaF  & $6.7(7)$ & $6.4(8)$ & $8(1)$ & $8.1(5)$ & $8(1)$\\
1M \ce{NH4Cl} & $6.5(6)$ & $6.4(3)$ & $6.7(8)$ & $6.4(8)$ & $6.6(8)$\\
0.5M \ce{NH4Cl} & $5.9(2)$ & $6.2(4)$ & $6.5(4)$ & $6.5(4)$ & $6.4(4)$\\
\bottomrule
\end{tabular}
\end{table}

\clearpage

\section{Temperature Dependence of Hydrogen Bond Jump Times}

The following figures include the average integrated hydrogen bond jump times, $\langle \tau_\mathrm{jump} \rangle$, across the range of temperatures tested for  supercooled liquid and interfacial systems.

\begin{figure}[h]
    \centering
    \includegraphics[width=\linewidth]{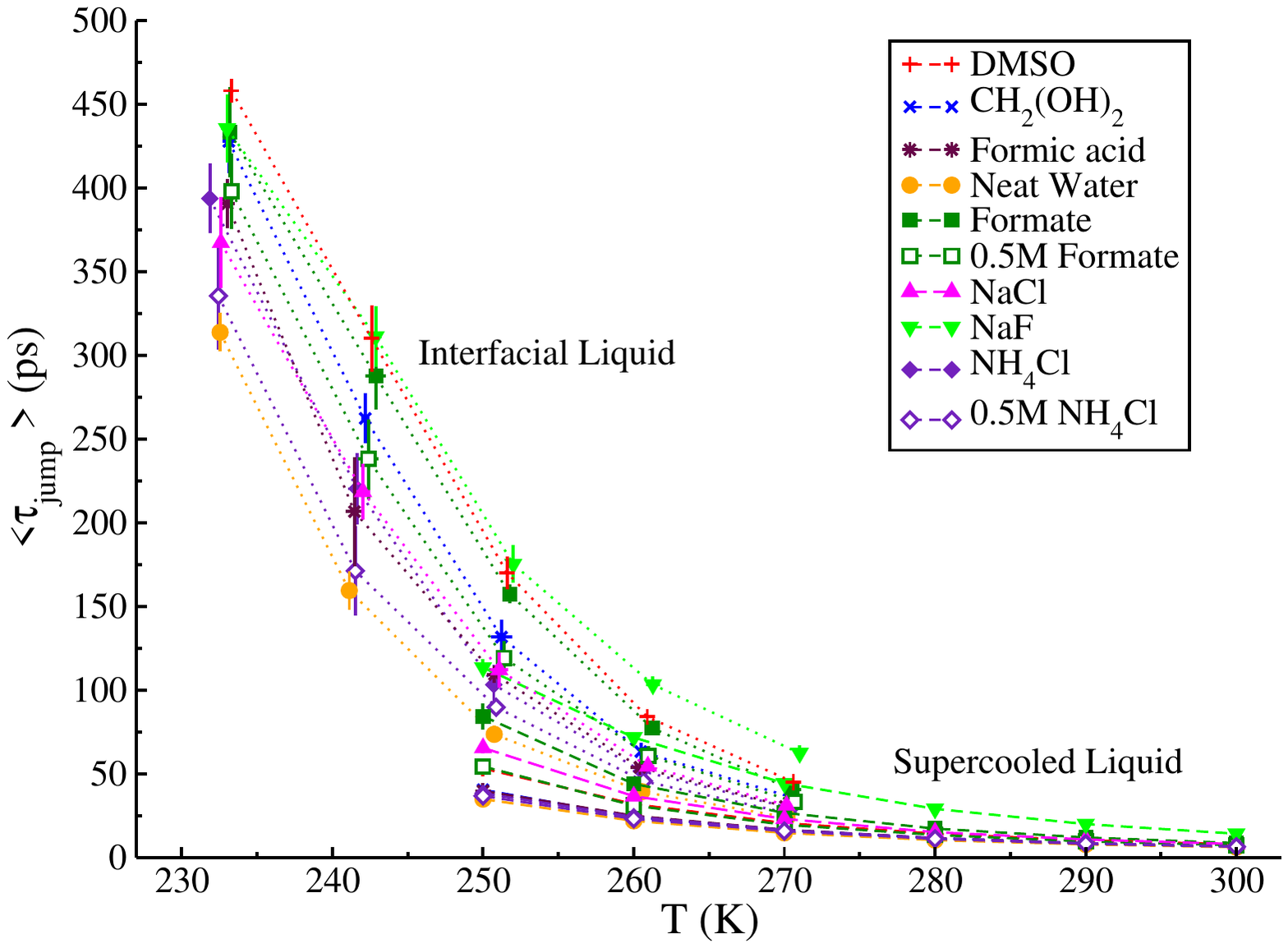}
    \caption{Average integrated hydrogen bond jump times (Eq. 10 in the main paper) as a function of temperature for bulk supercooled and confined interfacial liquids. In all cases, the presence of the ice crystal slows hydrogen  bonding dynamics relative to the bulk supercooled liquid. Molecular inclusions that are net acceptors generally correspond with slower hydrogen bond dynamics.}
    \label{fig:avgJumpTime}
\end{figure}

\begin{table}
\bibpunct{}{}{,}{n}{,}{,}
    \centering
    \caption{Intefacial widths (in \AA) calculated using the H-bond jump time as the order parameter. As was the case for the other widths, we do not observe a clear trend in temperature dependence, so an overall average was used in figure 5 of the main paper. CHOOH at 240K exhibited one particularly large outlier in the interfacial widths, leading to a confidence interval on the same order of magnitude as the reported width.}
\label{tab:TempjumpZd1090}
\begin{tabular}{ l | c c c c c c }
                \midrule
 
Solution & 230K & 240K & 250K & 260K & 270K\\
\midrule
Pristine & $11(5)$ & $12(4)$ & $13(2)$ & $10(2)$ & $9.8(8)$\\
1M DMSO &  $11(5)$ &  $9(2)$ &  $10(1)$ &  $8.7(6)$ &  $8.6(8)$\\
1M \ce{CH2(OH)2} & $9(2)$ & $12(4)$ & $9(2)$ & $10(2)$ & $9(1)$ \\
1M CHOOH & $10(2)$ & $10(10)$ & $10(4)$ & $9(2)$ & $9(1)$\\
1M \ce{HCOONa} & $12(5)$ & $12(5)$ & $11(2)$ & $12(2)$ & $11.0(7)$\\
0.5M \ce{HCOONa} & $10(2)$ & $14(4)$ & $14(4)$ & $10(2)$ & $10(1)$\\
1M NaCl & $8(2)$ & $9(2)$ & $9(3)$ & $10(2)$ & $10(2)$\\
1M NaF & $11(5)$ & $9(2)$ & $11(2)$ & $12(4)$ & $11(7)$\\
1M \ce{NH4Cl} & $9(2)$ & $10(8)$& $8.5(6)$ & $11(3)$ & $10(3)$\\
0.5M \ce{NH4Cl} & $12(5)$ & $15(5)$ & $10(1)$ & $10(2)$ & $10(1)$ \\
\bottomrule
\end{tabular}
\end{table}

\clearpage

\section{Fitting velocity profiles}
In order to calculate solid-liquid friction coefficients, $\kappa$
from Eq. (12) in the main text, the velocity profiles, $v_x(z)$,
obtained from each shearing simulation were fit assuming linear
behavior through each of the three regions of the simulation box; the
lower liquid, the solid, and the upper liquid. Parabolic functions
were designed to capture the negative slip behavior that links the
three regions,
\begin{equation}\label{vfit}
v(z) =
\begin{cases}
  v_{l} + m_{l}z & 0 \leq z < (z_{1} - w) \\
  v_{s} - \frac{1}{2}k(z-z_{1})^{2} & (z_{1}-w) \leq z < z_{1} \\
  v_{s}  & z_{1} \leq z < z_{2} \\
  v_{s} - \frac{1}{2}k(z-z_{2})^{2}  & z_{2} \leq z <( z_{2} + w)\\
  v_{s} - \frac{1}{2}kw^{2} - m_{l}(z-(z_{2} + w)) & (z_{2} + w) \leq z \\
\end{cases}
\end{equation}
Here, $v_{l}$ is the velocity of the liquid at the middle of the
liquid domain (the edge of the simulation box), and $v_{s}$ is the
velocity of the solid. The locations $z_{1}$ and $z_{2}$ are the edges
of the ice slab, and $w$ is the width of the interface (distinct from
$w_{10-90}$ mentioned in the main text). The parameter $m_{l}$ is the
slope of the velocity profile in the liquid regions of the box which
is related to the liquid-state viscosity, and 
\begin{equation}
    k = \frac{2 (v_{s} - v_{l} - m_{l} (z_{1} - w))}{w^2}~.
\end{equation}

Once the fits were obtained, the values for
$v_{x}(solid)$ and $v_{x}(liquid)$ for Eqs. (14) and (15) were sampled from the
fit. The $z$ locations used to sample the fit were determined by
structural measures. The $z$ location for $v_{x}(liquid)$ was taken to
be the Gibbs dividing surface of the interface, less half the 10$-$90 width
of the interface. Similarly, the $z$ location for $v_{x}(solid)$ was
taken to be the Gibbs dividing surface plus half the 10$-$90 width of the
interface.

\newpage

\subsection{Anion vs. Cation Uptake in the Interfacial Region}
\begin{figure}[h!]
    \centering
    \includegraphics[width=\linewidth]{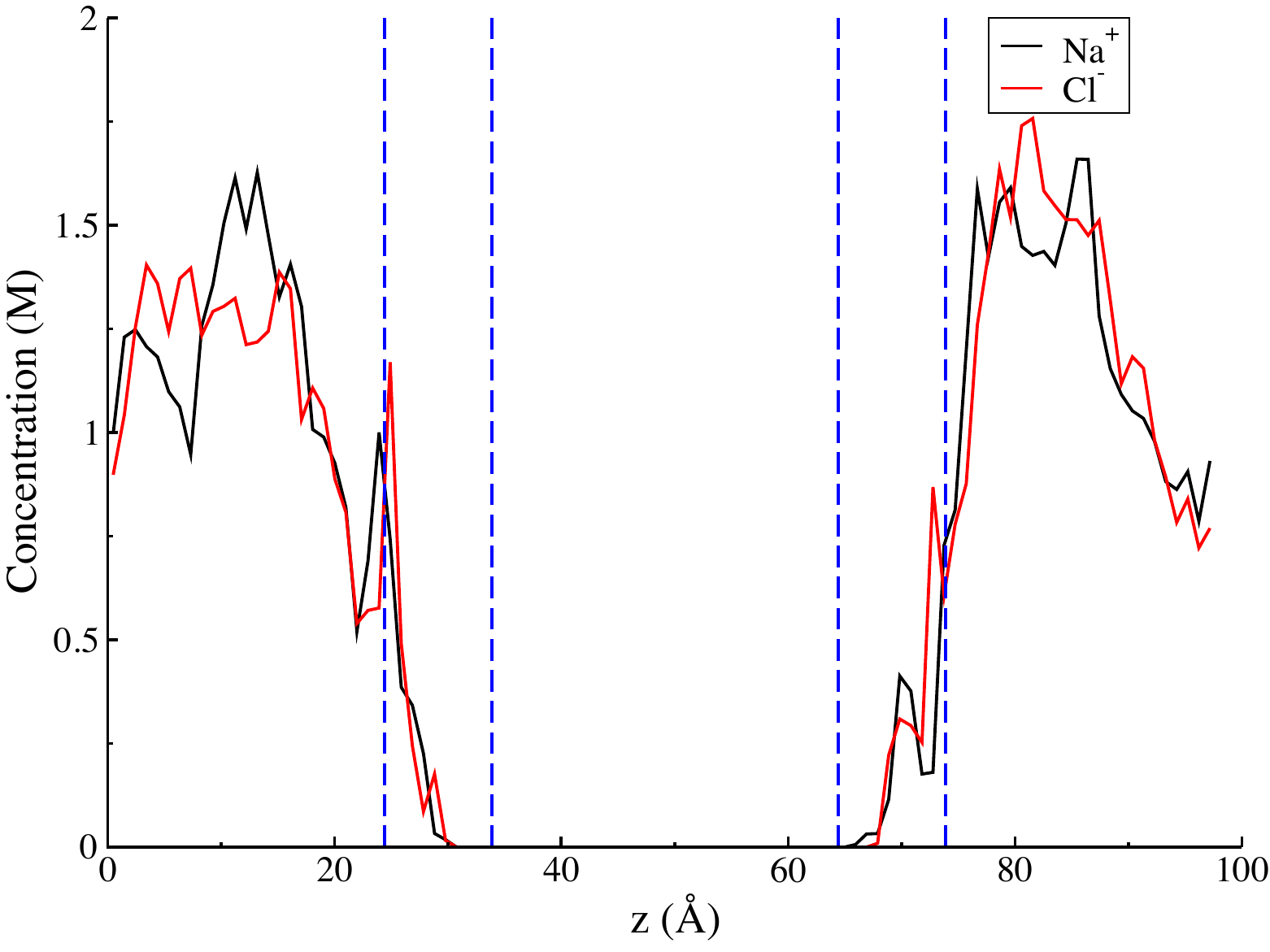}
    \caption{Average local \ce{Na+} and \ce{Cl-} concentrations for the five 1M NaCl solutions in contact with ice. The ice regions in the center of are bounded by four lines corresponding to the 90\% (inner) and 10\% (outer) bounds of the orientational order parameter, $q(z)$. No statistically significant differences between the two ion concentrations were observed.}
    \label{fig:NaCl_uptake}
\end{figure}

In the main text, we mentioned that anion and cation uptake in the interfacial regions were not appreciably different. In Fig. \ref{fig:NaCl_uptake} we show the sodium and chloride ion concentrations as a function of location in the simulation cell for the 1M \ce{NaCl} shearing simulations. The ions display local concentration profiles that are very similar, and the integrated uptake concentrations in the interfacial regions are also similar (0.20 M for \ce{Na+} vs. 0.24 M for \ce{Cl-}). Note that this difference is inside confidence intervals for 1M solutes, so we cannot distinguish between cation and anion uptake.

\bibliography{main}